\def\draftversion{false}
\RequirePackage{ifthen}
\ifthenelse{\equal{\draftversion}{true}}{
  \documentclass[aps,prb,10pt,galley,amsmath,amssymb,showpacs,
                 superscriptaddress]{revtex4}
}{
  \documentclass[aps,prb,10pt,twocolumn,showpacs,
                 superscriptaddress]{revtex4-1}
}

\usepackage{times}
\usepackage{amsmath}
\usepackage{amssymb}
\usepackage{graphicx}
\usepackage[usenames, dvipsnames]{color} % colors
\usepackage{bm} % bold math
\usepackage{subfigure}

\usepackage{ulem}  % for strike-out \sout

\def\Z2{$\mathbb{Z}_2$}
\def\I{\uppercase\expandafter{\romannumeral 1}}
\def\II{\uppercase\expandafter{\romannumeral 2}}
\def\III{{\uppercase\expandafter{\romannumeral 3}}}
\def\IV{{\uppercase\expandafter{\romannumeral 4}}}
\def\V{{\uppercase\expandafter{\romannumeral 5}}}

\def\bea{\begin{eqnarray}}
\def\nn{\nonumber\\}
\def\eea{\end{eqnarray}}
\def\beq{\begin{equation}}
\def\eeq{\end{equation}}
\def\eq#1{Eq.~(\ref{eq:#1})}

\def\ket#1{\vert#1\rangle}

\def\ip#1#2{\langle#1\vert#2\rangle}
\def\me#1#2#3{\langle#1\vert#2\vert#3\rangle}

\def\dag{^\dagger}

\def\wt#1{\widetilde{#1}}

\def\O{\Omega}
\def\Ot{\wt{\O}}

\def\BK{_\textrm{BK}}
\def\GD{_\textrm{GD}}
\def\VL{_\textrm{VL}}

\def\veps{\varepsilon}

\def\k{{\bf k}}
\def\bnk{_{n\k}}
\def\bmk{_{m\k}}

\def\lam{\lambda}
\def\ol{(\lam)}
\def\bl{_\lam}

\def\bb{^{(0)}}
\def\tt{^{(1)}}
\def\ub{u\bb}
\def\Ab{A\bb}
\def\ut{u\tt}
\def\At{A\tt}

\def\Ao{\Ab}

\def\para{\partial_a}
\def\parb{\partial_b}
\def\parx{\partial_x}
\def\pary{\partial_y}

\def\parl{\partial_\lam}
\def\part{\wt{\partial}}

\def\Tr{{\rm Tr\,}}

\def\fpii{\frac{1}{4\pi}}

\def\Wd{W^\dagger}

\def\Bbar{\overline{B}}
\def\Bbbar{\overline{\overline{B}}}
\def\cxy{{[x,y]}}

\def\cS{\mathcal{S}}

\def\calA{{\cal A}}
\def\Z2{$\mathbb{Z}_2$}

\def\bea{\begin{eqnarray}}
\def\nn{\nonumber\\}
\def\eea{\end{eqnarray}}
\def\beq{\begin{equation}}
\def\eeq{\end{equation}}
\def\eq#1{Eq.~(\ref{eq:#1})}

\def\ket#1{\vert#1\rangle}

\def\ip#1#2{\langle#1\vert#2\rangle}
\def\me#1#2#3{\langle#1\vert#2\vert#3\rangle}

\def\dag{^\dagger}

\def\wt#1{\widetilde{#1}}

\def\O{\Omega}
\def\Ot{\wt{\O}}
\def\D{\Delta}

\def\BK{_\textrm{BK}}
\def\GD{_\textrm{GD}}
\def\VL{_\textrm{VL}}

\def\veps{\varepsilon}

\def\k{{\bf k}}
\def\bnk{_{n\k}}
\def\bmk{_{m\k}}
\def\r{\bf r}

\def\lam{\lambda}
\def\ol{(\lam)}
\def\bl{_\lam}

\def\bb{^{(0)}}
\def\tt{^{(1)}}
\def\ub{u\bb}
\def\Ab{A\bb}
\def\ut{u\tt}
\def\At{A\tt}

\def\Ao{\Ab}

\def\para{\partial_a}
\def\parb{\partial_b}
\def\parx{\partial_x}
\def\pary{\partial_y}

\def\parl{\partial_\lam}
\def\part{\wt{\partial}}

\def\Tr{{\rm Tr\,}}

\def\cT{\mathcal{T}}
\def\cP{\mathcal{P}}

\def\Green#1{\textcolor{OliveGreen}{#1}}

\begin{document}

\title{Gauge-discontinuity contributions to Chern-Simons orbital magnetoelectric coupling}
\date{today}

\author{Jianpeng Liu}

\affiliation{ Kavli Institute for Theoretical Physics, University of California, Santa Barbara,
CA 93106, USA } 
\affiliation{ Department of Physics and Astronomy, Rutgers University,
 Piscataway, NJ 08854-8019, USA }
 
\author{David Vanderbilt}

\affiliation{ Department of Physics and Astronomy, Rutgers University,
 Piscataway, NJ 08854-8019, USA }

\date{\today}

\begin{abstract}
We propose a new method for calculating
the Chern-Simons orbital magnetoelectric coupling,
conventionally parametrized in terms of a phase angle $\theta$.
According to previous theories,
$\theta$ can be expressed as a 3D Brillouin-zone
integral of the Chern-Simons 3-form defined in terms
of the occupied Bloch functions.  Such an expression
is valid only if a smooth and periodic gauge has been
chosen in the entire Brillouin zone,  and even then,
convergence with respect to the $\k$-space mesh
density can be difficult to obtain.
In order to solve this problem,
we propose to relax the periodicity condition in
one  direction (say, the $k_z$ direction)
so that a gauge discontinuity is introduced on
a 2D $\mathbf{k}$ plane normal to $k_z$.
The total $\theta$ response then has contributions
from both the integral of the Chern-Simons 3-form over the 3D bulk BZ
and the gauge discontinuity expressed as a 2D integral over the
$\mathbf{k}$ plane. Sometimes the boundary plane
may be further divided into subregions by 1D ``vortex loops" which
make a third kind of contribution to the total $\theta$,
expressed as a combination of Berry phases around the vortex loops.
The total $\theta$ thus consists of three terms which
can be expressed as integrals over 3D, 2D  and 1D manifolds. When
time-reversal symmetry is present and the gauge in the bulk BZ is
chosen to respect this symmetry, both the 3D and 2D integrals
vanish; the entire contribution then comes from the vortex-loop integral,
which is either 0 or $\pi$ corresponding to the $\mathbb{Z}_2$
classification of 3D time-reversal invariant insulators.
We demonstrate our method by applying it to the Fu-Kane-Mele
model with an applied staggered Zeeman field.
\end{abstract}


\maketitle

%%%%%%%%%%%%%%%%%%%%%%%%%%%%%%%%%%%
%\marginparwidth 2.7in
%\marginparsep 0.5in
%\def\dvm#1{\marginpar{\small DV: #1}}
%\def\jpm#1{\marginpar{\small JPL: #1}}
%\def\scr{\scriptsize}
%%%%%%%%%%%%%%%%%%%%%%%%%%%%%%%%%%%

%-------- MARGIN COMMENTS --------------
\def\scr{\scriptsize}
\ifthenelse{\equal{\draftversion}{true}}{
  \marginparwidth 2.7in
  \marginparsep 0.5in
  \newcounter{comm} % counter for commentaries
  % increase counter
  \def\commnext{\stepcounter{comm}}
  % commentary in text
  \def\commtext{{\bf\color{blue}[\arabic{comm}]}}
  % commentary in margin
  \def\commmar{{\bf\color{blue}[\arabic{comm}]}}
  % comment commands for all authors
  \def\dvm#1{\commnext\marginpar{\small DV\commmar: #1}\commtext}
  \def\jlm#1{\commnext\marginpar{\small JPL\commmar: #1}\commtext}
  \def\mlab#1{\marginpar{\small\bf #1}}
  \def\tnewpage{\newpage\marginpar{\small Temporary newpage}}
  \def\tfootnote#1{\Green{\scr [FOOTNOTE: #1]}}
}{
  \def\dvm#1{}
  \def\jlm#1{}
  \def\mlab#1{}
  \def\tnewpage{}
  \def\tfootnote#1{\footnote{#1}}
}
%----------------------------------

%--------------------------------------------------
\section{Introduction}
%---------------------------------------------------

Magnetoelectric coupling is an interesting but complicated phenomenon
that can occur in some insulating solids when
an electric polarization $\mathbf{P}$
is linearly induced by an external magnetic field $\mathbf{B}$,
or conversely, when a magnetization $\mathbf{M}$ is generated by an
applied electric field $\mathbf{E}$.
The linear ME coupling coefficient is a rank-2 tensor
defined as
\beq
\alpha_{ab}=\frac{\partial M_b}{\partial E_a}{\Big\vert}_{\mathbf{E}=0}
=\frac{\partial P_a}{\partial B_b}\Big\vert_{\mathbf{B}=0}
\label{eq:me-def}
\eeq
where $a$, $b=\{x, y, z\}$ denote the directions in real space.
ME phenomena have contributions from both electronic and
lattice degrees of freedom, where
the electronic contribution refers to the ME response when the
ions are completely frozen, while the lattice contribution takes into account
the response that is mediated by ionic
displacements. Moreover, depending
on the origin of the  $\mathbf{E}$-induced magnetization,
each of the two contributions can be further decomposed into
spin and orbital components.\cite{fiebig2005review, malashevich2010ome}

The spin contribution to the ME response (from both electronic and
lattice degrees of freedom) has been thoroughly studied with
well established theoretical methods in typical magnetoelectrics
such as Cr$_2$O$_3$.\cite{cr2o3-lattice,spaldin-prl11,malashevich-prb12,ye-prb14}
On the other hand, the orbital ME response is theoretically
more challenging and intriguing. It has been shown
that the frozen-ion orbital ME coupling consists of two terms.
One term can be expressed as a standard linear
response of the Bloch functions to external electric or magnetic fields,
denoted as the ``Kubo term'',  while the other, known as the
Chern-Simons term, is isotropic and is completely determined
by the unperturbed ground-state
wavefunctions.\cite{malashevich2010ome,essin-prb10}

The Chern-Simons orbital ME coupling has drawn
significant attention recently due to the interest in topological
phases in condensed-matter physics.
Not surprisingly,
in the presence of either time-reversal ($\cT$) or inversion
($\cP$) symmetry, the ME responses coming from the spin terms
and from the Kubo-like orbital terms all vanish.  However,
there can still be an exotic isotropic ME response, which vanishes
in an ordinary insulator but takes values of $\pm e^2/2h$ in
$\cT$-respecting strong topological insulators\cite{kane-rmp10,
zhang-rmp11} and in $\cP$-respecting axion
insulators,\cite{hughes-prb11,turner-prb12} arising from the
Chern-Simons term.\cite{qi-prb08,essin-prl09,turner-prb12}

This Chern-Simons coupling is conventionally parametrized by a
dimensionless phase angle $\theta$ via
\beq
\alpha^{\textrm{CS}}_{ab}=\frac{\theta e^2}{2\pi h}\delta_{ab}\;,
\eeq
where $\theta$ is expressed as an
integral of the Chern-Simons 3-form over
the 3D Brillouin zone (BZ),
\beq
\theta = -\fpii\int d^3k \; \veps_{abc} \, \Tr{[A_a \parb A_c
    -\frac{2}{3}iA_aA_bA_c]} \;.
\label{eq:theta-a}
\eeq
Here $A_a$, $A_b$ and $A_c$ are the Berry connection matrices of the occupied
Bloch bands, and the trace is taken over the occupied bands
(see Sec.~\ref{sec:def}).
For TIs and axion insulators, $\theta=\pm\pi$. In the more
general cases that $\cT$ and  $\cP$ are both broken,
$\theta$ is no longer quantized as $\pm\pi$,
and other components of the ME response contribute as well.

The Chern-Simons ME coupling has several interesting properties.
First, a material with a non-zero Chern-Simons
ME coupling can be considered as a medium exhibiting axion
electrodynamics,\cite{axion-ed} where an additional
term $\Delta\mathcal{L}=\alpha^{\textrm{CS}}\mathbf{E}\cdot\mathbf{B}$
is added to the conventional Lagrangian of electromagnetic fields
in media. The electrodynamics with such an axion coupling
turns out to be invariant under $\theta\to\theta+2\pi$.\cite{axion-ed}

Secondly, $\theta$ is physically measurable only if it
varies in space or time.\cite{essin-prb10} In particular,
for a time-independent crystal with a surface truncation,
the presence of the bulk Chern-Simons coupling manifests itself
as a surface anomalous Hall effect, where the anomalous
Hall conductance is proportional to $\theta$ through
$\sigma_{xy}=\theta e^2 /(2\pi h)$.
The connection between the surface anomalous Hall effect and
the bulk Chern-Simons ME coupling provides an intuitive
explanation of the ambiguity of $\theta$ as follows.
Suppose an insulating quantum anomalous Hall (QAH) layer with non-zero
Chern number $C$ is
wrapped around a 3D crystallite having an original bulk
value of $\theta$, such that it interacts only weakly with all
of the surfaces.  Then
the the new surface anomalous Hall conductance would be
$\sigma_{xy}=\theta e^2 /(2\pi h)+C e^2/h$, which we can
be interpreted as a change $\theta\rightarrow\theta+2\pi C$.
Thus, such a freedom to coat the surfaces with Chern layers
implies the need for a $2\pi$ ambiguity in defining $\theta$.
The ambiguity in $\theta$
is closely analogous to the ambiguity in the definition of
the bulk electric polarization, which can be regarded as being
due to the freedom of adding or removing an integer number
of charges per surface unit cell, as by filling or emptying
a surface band.\cite{vanderbilt-prb93}

Despite these intriguing properties,
up to now it has remained challenging to
calculate $\theta$ accurately using \eq{theta-a} for many systems
of interest.  For example, as reported in Ref.~\onlinecite{coh-prb11},
the calculated $\theta$ on an 11$\times$11$\times$11 first-principles \k\ mesh
for Bi$_2$Se$_3$, one of the prototype TIs,
is only $\sim 35\%$ of $\pi$. Similarly, in Ref.~\onlinecite{essin-prl09},
the authors calculated the ME response of
the Fu-Kane-Mele model with applied staggered Zeeman field.
As the system approaches the TI phase, however,
the authors switched to some indirect methods to
compute $\theta$, because a direct numerical implementation of
\eq{theta-a} became difficult to converge.
In other words, despite its theoretical importance, \eq{theta-a}
has not been straightforward to calculate in practice.

The essential problem is that the integrand in \eq{theta-a}
is gauge-dependent. As a result, in order to implement \eq{theta-a}
numerically on a discrete \k\ mesh, one has to adopt a smooth
and periodic gauge over the entire 3D BZ. On the other hand, as is
well known,  nontrivial topological indices usually bring
some obstructions against constructing a smooth and periodic gauge in the BZ.
For example, for a 2D quantum anomalous Hall (QAH) insulator (such as
the Haldane model\cite{Haldane-model}) with non-zero Chern number,
it is simply impossible to construct a smooth and periodic gauge
in the entire 2D BZ.  This implies that \eq{theta-a} would completely
break down for a 3D analogue of a 2D QAH insulator,
%%%%%%%%%%%%
\footnote{A 3D QAH insulator is defined as a 3D insulator
with the property that for at least one orientation of 2D $\k$ slices
through the BZ, the Chern number of these slices is non-zero.
Such a system is adiabatically connected to one made by
stacking QAH layers in the third spatial dimension.}
%%%%%%%%%%%%%
so we regard these cases as beyond the scope of the present work.
For 2D and 3D \Z2\ TIs, it is impossible to construct a smooth
and periodic gauge respecting $\cT$ symmetry
throughout the BZ,\cite{fu-prb06,soluyanov-prb12} although
in principle a smooth and periodic gauge breaking $\cT$
symmetry is allowed.\cite{soluyanov-prb12}
As a result, for \Z2\ TIs (and for $\cT$-broken systems close to a
\Z2-odd\ phase) the constraint of being both smooth and periodic
is typically too strong, forcing the gauge to be strongly twisted
in the BZ to satisfy both conditions.
This makes the numeric implementation of \eq{theta-a} difficult.

In this paper we propose a new method to compute the Chern-Simons
orbital ME coefficient. The general idea is to relax the
periodicity condition on the gauge in one direction, say the $k_z$ direction,
thus introducing some gauge discontinuity on a 2D $\k$ plane (normal
to $k_z$), denoted by $\mathcal{S}$.
Then the total $\theta$ has one contribution from the bulk-BZ integral
of \eq{theta-a} plus a second one arising from the gauge discontinuity.
Furthermore, as will be shown
in Sec.~\ref{sec:vl}, $\mathcal{S}$ may also
be divided into subregions by 1D ``vortex loops" (Sec.~\ref{sec:vldef}),
each of which makes a contribution to the total $\theta$ in the form of an
average of two Berry phases computed around the loop.
The total $\theta$ can then be expressed as the sum of
the 3D integral over the bulk BZ ($\theta\BK$),
the 2D integral over the gauge-discontinuity plane ($\theta\GD$),
and the 1D integral(s) over the vortex loop(s) ($\theta\VL$).

This method can be generalized to situations where
the BZ is divided into multiple subvolumes, with these
subvolumes meeting at multiple 2D surface patches where the
gauge discontinuities reside.
Furthermore, the 2D surface patches may meet at some 1D curves,
which again have to be treated as as vortex lines in general.
And again, the subvolumes, surface patches, and vortex lines
all make contributions to the total $\theta$. However, the
definition of a vortex line becomes trickier in this more
generalized case, which we therefore leave for future study.

The advantage of our method is that the gauge can be made smoother
in the bulk BZ because the periodicity condition is relaxed, so
that it becomes much easier to get numeric convergence using
\eq{theta-a}. The loss of periodicity is then compensated
by contributions from the gauge discontinuities, and possibly
from vortex loops as well.
We will show that the formulas for the gauge discontinuity and vortex
terms take simple forms and can be implemented efficiently
in practical numerical calculations.

This paper is organized as follows. In Sec.~\ref{sec:pre} we
review the definitions of the Berry connection and curvature
and introduce the bulk formula for $\theta$. We also put the main
idea into a more specific context and make a formal statement
of the problem. In Sec.~\ref{sec:surf} we derive a formula for
$\theta\GD$, which is expressed as a 2D integral over
the boundary where the gauge discontinuity resides, and
discuss the properties of this formula. In Sec.~\ref{sec:vl} we
discuss why the vortex-loop term
is needed and derive a formula for it. We also
show that the quantized $\theta$ in TIs is completely
determined by the vortex-loop term when a $\cT$-symmetric gauge is
chosen in the bulk BZ. In Sec.~\ref{sec:apply}, we demonstrate the
method by applying it to the Fu-Kane-Mele model with a staggered
Zeeman field.  Finally, we summarize in Sec.~\ref{sec:summary}.

%------------------------------------------------------------
\section{Preliminaries}
\label{sec:pre}
%------------------------------------------------------------

In this section, we first review the definitions of
some basic quantities, such as Berry curvatures and
Berry connections, that will be used frequently in the paper.
We also rewrite the bulk formula for $\theta$, \eq{theta-a},
in a more explicit form. Finally we explain
the main idea in more detail and make a formal statement
of the problem and the goals.

%------------------------------------------------------------
\subsection{Definitions}
\label{sec:def}
%------------------------------------------------------------

We adopt the following definitions.  The Berry connection matrix is
\beq
A_{a,mn}(\k)=i\me{u\bmk}{\para}{u\bnk} \;,
\label{eq:Adef}
\eeq
where $u\bnk({\r})=e^{-i\k\cdot\r}\psi\bnk(\r)$ are the cell-periodic
Bloch functions, and
$a$ and $b$ run over the three primitive reciprocal
lattice directions with $\partial_a\equiv\partial/\partial k_a$.
Indices $m$ and $n$ run over the occupied Block bands,
possibly after the application of a gauge transformation
$U_{nm}(\k)$ to smoothen them in $\k$-space.
The wavevector components $k_x$ etc.\ are rescaled to run over
$[0, 2\pi]$, and correspondingly the real-space coordinates
$x$ etc.\ run over $[0,1]$.
We shall start dropping the explicit $\k$ arguments and
subscripts, keeping in mind that everything is a function of $\k$.
Then the non-covariant Berry curvature tensor is
\beq
\O_{ab,mn}= %= \para A_{b,mn} - \parb A_{a,mn} =
i\, \ip{\para u_{m}}{\parb u_{n}} -i\, \ip{\parb u_{m}}{\para u_{n}}
\;,
\label{eq:Odef}
\eeq
while
\beq
\Ot_{ab,mn} = \O_{ab,mn} -i{[A_a,A_b]}_{mn}
\label{eq:Otdef}
\eeq
is the covariant one (that is, unlike $\O_{ab,mn}$, it transforms
in the standard way under a gauge transformation).

The Chern-Simons  coupling $\theta$ has been defined in \eq{theta-a},
where the trace is over the occupied band indices.
Using the cyclic property of the trace,
\eq{theta-a} can be written in the more explicit form
\beq
\theta = -\fpii\int d^3k \;  \Tr\Big[
   A_x \O_{yz} + A_y \O_{zx} + A_z \O_{xy} -2i [A_x,A_y]A_z \Big] \;.
\label{eq:theta-b}
\eeq
We can also choose to replace one of the non-covariant Berry curvatures
with a covariant one to get
\beq
\theta = -\fpii\int d^3k \;  \Tr\Big[
A_x \O_{yz} + A_y \O_{zx} + A_z \Ot_{xy} -i [A_x,A_y]A_z \Big] \;,
\label{eq:theta-d}
\eeq
which turns out to be convenient for the derivation
of $\theta\GD$ as will be shown in Sec.~\ref{sec:surf}.

%------------------------------------------------------------
\subsection{Statement of the problem}
\label{sec:statement}
%------------------------------------------------------------

Assume that the gauge has been chosen such that it is smooth
and periodic in the $k_x$ and $k_y$ directions and smooth in
$k_z\in[-\pi,\pi]$, but not periodic in $k_z$.  (The
$k_z$ location of the boundary can easily be generalized.) From
now on $\k=(k_x,k_y)$ denotes a point in the 2D slice at
$k_z=\pm\pi$, and $\ket{\ub}$ and $\ket{\ut}$ denote the wavefunctions
just below and above the discontinuity plane respectively.
For this reason we refer to $\ket{\ub}$ and $\ket{\ut}$ as associated
with the ``bottom'' and ``top'' planes, even though these are obtained
from the top and bottom of the original BZ, respectively.
The corresponding Berry potentials are
$\Ab_x$ and $\Ab_y$ on the bottom plane and
$\At_x$ and $\At_y$ on the top plane.
The states at the top and bottom are physically identical, so we can
define a unitary matrix $U(\k)$ relating them via
\beq
\ket{\psi^{(1)}\bmk}=\sum_n \ket{\psi^{(0)}\bnk}\,U_{nm}(\k)
\label{eq:statement-bloch}
\eeq
for the original Bloch functions or
\beq
\ket{\ut\bmk} = e^{i 2\pi z}\sum_n \ket{\ub\bnk}\,U_{nm}(\k)
\label{eq:statement}
\eeq
for the cell-periodic Bloch functions.
Our goal is to calculate the contribution $\theta\GD$ coming from
this gauge discontinuity, such that if we add this contribution to
the  bulk volume integral $\theta\BK$ as in \eq{theta-b},
we get the correct total $\theta$.
Later, we shall see that there may also
be a contribution $\theta\VL$ from vortex loops around which the
gauge discontinuity circulates by an integer multiple of $2\pi$,
so that the total axion coupling is given by
\beq
\theta=\theta\BK+\theta\GD+\theta\VL \;,
\label{eq:thetatot}
\eeq
i.e., a sum of contributions evaluated on 3D, 2D, and 1D manifolds.

%------------------------------------------------------------
\section{Calculation of $\theta\GD$ on a planar surface}
\label{sec:surf}
%------------------------------------------------------------

\begin{figure}
\centering
\includegraphics[width=8.8cm]{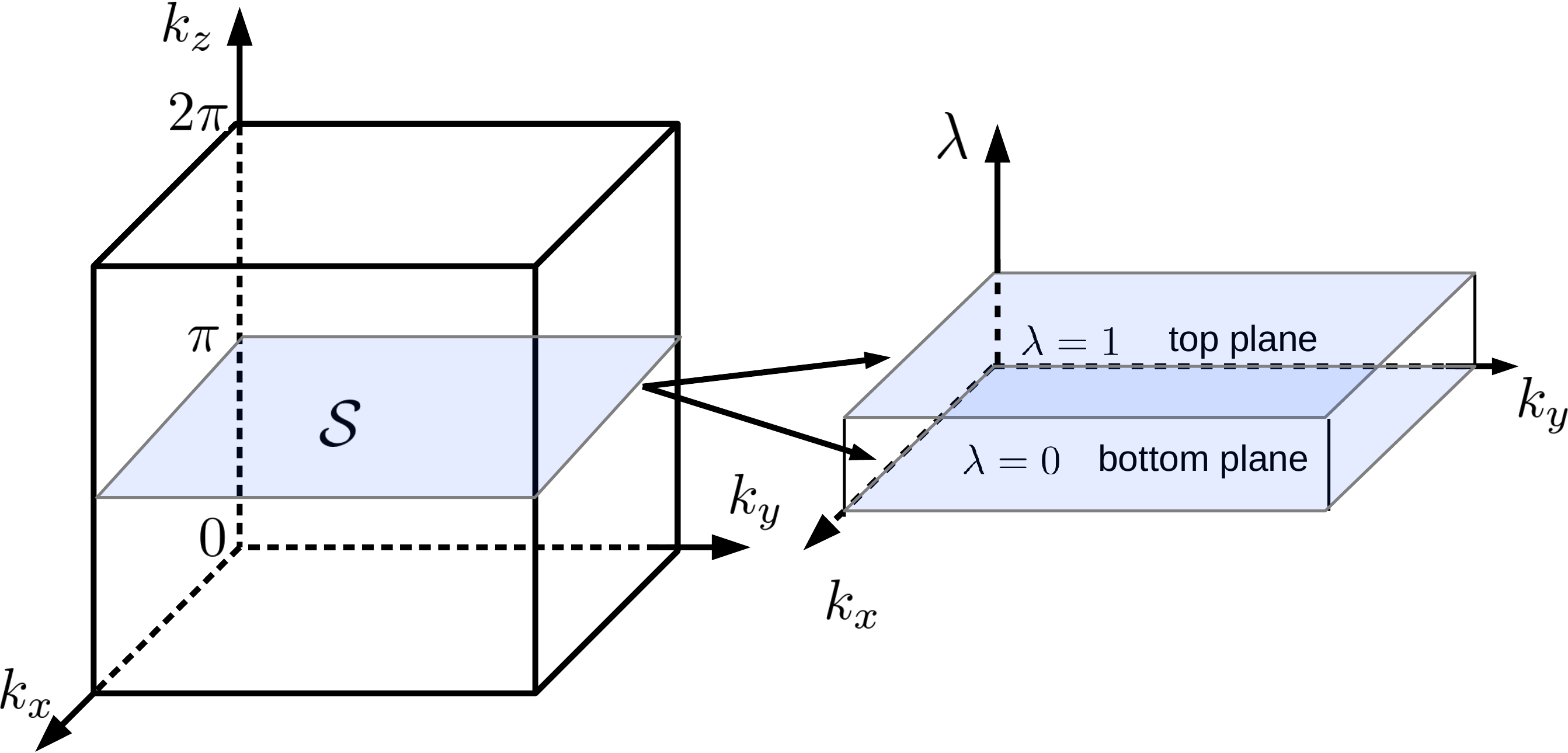}
\caption{\label{fig:gd}
A planar gauge discontinuity $\cS$ in the 3D BZ can be expanded into a
fictitious slab whose thickness dimension is described by a
parameter $\lambda\in [0,1]$ that interpolates smoothly
between the gauge just below ($\lambda\!=\!0$)
and just above ($\lambda\!=\!1$) the plane $\cS$.}
\end{figure}

In this section, we derive a formula for $\theta\GD$ and
discuss various properties of the formula.
We assume, as above, that the gauge discontinuity occurs on
the $k_z=\pm\pi$ plane as schematically shown in Fig.~\ref{fig:gd},
and is described by the unitary matrices
$U_{\k}$ as a function of $\k$ lying in the 2D plane.  We let
\beq
U(\k)=e^{-iB(\k)}
\label{eq:U}
\eeq
where $B(\k)$ is a Hermitian matrix that varies smoothly with $\k$
in the 2D plane.  Note that $B(\k)$ is basically just $i\ln(U(\k))$, but
a set of branch choices is involved in picking a particular $B$.
That is, in the representation that diagonalizes $B$, we can add
$2\pi n_j$ to the $j$'th eigenvalue without changing $U$
($n_j$ is an arbitrary integer).  For now we insist that
the branch choice is made in such a way that
$B(\k)$ is continuous, with no $2\pi$ discontinuities in any of
its eigenvalues throughout the 2D \k\ plane,
but this condition will be relaxed in Sec.~\ref{sec:vl}.

% ::::::::::::::::::::::::::::::::::::::::::::::::::::::::::::
\subsection{Formalism}
\label{sec:formalism}
% ::::::::::::::::::::::::::::::::::::::::::::::::::::::::::::

Our strategy is to introduce a parameter $\lambda$ and define
$\ket{\psi_{m\k}(\lam)}$
in such a way
that it smoothly interpolates from one gauge to the other
as shown in Fig.~\ref{fig:gd}, i.e.,
\beq
%\ket{u_{m\k}\ol} = e^{-i2\pi z}\sum_n \ket{\ub_{n\k}} \, W_{nm}(\k,\lambda)
\ket{\psi_{m\k}\ol} = \sum_n \ket{\psi^{(0)}_{n\k}} \, W_{nm}(\k,\lambda)
\label{eq:interpolation}
\eeq
where
\beq
W(\k,\lambda) = e^{-i\lam B(\k)}
\label{eq:Wdef}
\eeq
where $W(\k,\lambda)$ is a unitary matrix defined
so that $W(\k , 0)=1$ and $W(\k , 1)=U(\k)$. Note
that $W(\k,\lambda)$ commutes with $B(\k)$.
We shall again begin
dropping the $\k$ labels, and will frequently use  $W$ and $B$ below.

We then calculate the gauge-discontinuity contribution to $\theta$,
denoted by $\theta\GD$, by integrating \eq{theta-d} over
the region $\lam\in[0,1]$, where \eq{theta-d} is applied in
$(k_x,k_y,\lam)$ space instead of $(k_x,k_y,k_z)$ space.
A straightforward set of calculations shows that
the Berry connections in the
$k_x$, $k_y$, and $\lambda$ directions are respectively
\bea
&& A_x\ol = \Wd\ol\,\Ao_x\,W\ol + i \Wd\ol\,\parx W\ol \;,
\label{eq:Ax}\\
&& A_y\ol = \Wd\ol\,\Ao_y\,W\ol + i \Wd\ol\,\pary W\ol \;,
\label{eq:Ay}\\
&&A\bl\ol = B \;,
\label{eq:A_l}
\eea
where $A_{x(y)}^{(0)}$ is the Berry connection evaluated at the bottom plane
as defined earlier.  We also write
\beq
\theta\GD = -\fpii\int d^2k \;  G(\k) \; ,
\label{eq:tGD}
\eeq
where
\beq
G = \int_0^1 d\lam\;  \Tr\Big[
  A_x \O_{y\lam} - A_y \O_{x\lam} + A_\lam \Ot_{xy}
   -i [A_x,A_y]A_\lam \Big]
\label{eq:G}
\eeq
is the contribution from a particular $\k$.
Then $G$ can be written as the
sum of three contributions, $G=G_1+G_2+G_3$, where
\bea
&&G_1(\k)=\int_0^1 d\lam\; \Tr[B\, \Ot_{xy}] \;,\;
\label{eq:G1} \\
&&G_2(\k)=\int_0^1 d\lam\; \Tr[A_x \O_{y\lam} - A_y \O_{x\lam}] \;,\;
\label{eq:G2} \\
&&G_3(\k)=\int_0^1 d\lam\; \Tr\Big[-i\,[A_x,A_y] \, B \Big] \;.
\label{eq:G3}
\eea

The $G_1$ term is easily evaluated. Because $\Ot_{xy}$ is gauge-covariant, it
follows that $\Ot_{xy}\ol=\Wd\ol\,\Ot\bb_{xy}\,W\ol$. But
$[B,W\ol]=0$, so that the integrand is independent of $\lam$, and it
follows that
\beq
G_1(\k)=\Tr[B\,\Ot^{(0)}_{xy} ]
\label{eq:G1final} \;.
\eeq
Here no $\lambda$ integration is needed.

In order to evaluate $G_2$ and $G_3$, we need to evaluate objects
such as $\parx W(\lam)$ in \eq{Ax}, which can be done
by noting that the derivative of an exponential of a matrix can
be written as
\beq
\parx{e^{-i\lam M}}=
-i\,\int_0^\lam d\mu\; e^{-i\,\mu M}(\parx M)e^{-i\,(\lam-\mu) M} \;.
\label{eq:partial_exp}
\eeq
This motivates us to define
\beq
\Bbar_{a}\ol=\int_0^\lam d\mu\;e^{-i\mu B}\,B_{a}\,e^{i\mu B} \;,
\label{eq:Bbarxy}
\eeq
where $B_{a}\equiv\partial_a B$. Then \eq{Wdef} gives
\beq
\partial_a W\ol=\partial_a{e^{-i\lam B}}=
-i\Bbar_a\ol W\ol
\label{eq:Wx}
\eeq
where $a=\{x, y\}$, and Eqs.~(\ref{eq:Ax}-\ref{eq:Ay}) become
\beq
A_{a}\ol\, = \Wd\,\calA_{a}\, W \; ,
\label{eq:Alxy}
\eeq
where
\beq
\calA_{a} = \Ao_{a} +  \Bbar_{a} \; .
\label{eq:cAxy}
\eeq
The dependence on $\lam$ is implicit.

Now for the $G_2$ term we need to compute terms like $\parl A_x$.
Using \eq{Alxy} and $\partial_{\lam}\,W(\lam)=-iB W(\lam)$, it becomes
\beq
\parl A_x \,
= i\Wd\,[B,\calA_x]\, W + B_x \; .
\label{eq:parl_Ax}
\eeq
Recalling that $\O_{x\lam}=\parx A_\lam- \parl A_x$
and $\parx A_\lam=B_x$,
we get a nice cancellation, and can write
\beq
\O_{x\lam} = -i\,\Wd\,[B,\calA_{x}]\,W \; ,\;\nn
\O_{y\lam} = -i\,\Wd\,[B,\calA_{y}]\,W \;.
\label{eq:Ol}
\eeq
Substituting these expressions into Eq.~(\ref{eq:G2}) then gives
\bea
G_2(\k) =\int_0^1 d\lam\; \Tr\Big[
 2i\,B\,[\calA_x,\calA_y] \Big] \; .
\label{eq:G2final}
\eea
As it happens, this is
almost the same as the expression for
$G_3$ in \eq{G3}.  Since $B$ commutes with
$W$, we can use the representation-invariance and cyclic properties
of the trace to write it as
\beq
G_3(\k)=\int_0^1 d\lam\; \Tr\Big[-i\,B\,[\calA_x,\calA_y] \Big] \;.
\label{eq:G3final}
\eeq
Thus, this term cancels half of $G_2$.

Restoring the explicit $\lambda$ dependencies, we get
\beq
G= \Tr\Big[ B \, \Big( \Ot\bb_{xy}
    +i\int_0^1 d\lam\,[\calA_x\ol,\calA_y\ol] \Big) \Big]\;,
\label{eq:Gcurly}
\eeq
which is a remarkably simple result in the end.
Using \eq{cAxy}, this can be written
explicitly as
\beq
G(\k)=\Tr\Big[B\Big( \O^{(0)}_{xy}+\Bbbar_\cxy
+i[\,\Bbbar_x,\Ao_y\,] -i[\,\Bbbar_y,\Ao_x\,] \Big) \Big]\; ,
\label{eq:Gfull}
\eeq
where
\bea
&& \Bbbar_x = \int_0^1 d\lam\;\Bbar_x(\lam) \;,
\label{eq:bbx}\\
&& \Bbbar_y = \int_0^1 d\lam\;\Bbar_y(\lam) \;,
\label{eq:bby}\\
&& \Bbbar_\cxy = i\int_0^1 d\lam\;
         [\,\Bbar_x(\lam),\Bbar_y(\lam)\,]
\label{eq:bbxy}\;.
\eea
%
%The three double-barred quantities depend only on $B$, $B_x$, and $B_y$.
Eq.~(\ref{eq:Gfull}) is one of the central results of this paper.

We would like to make some remarks on the formula for $\theta\GD$.
First, the results are almost independent of the actual states
at the top and bottom of the gauge discontinuity.  The only way these
come in is through the Berry potentials $\Ao_x$ and $\Ao_y$ and
the Berry curvature $\O^{(0)}_{xy}$ defined on one of the planes.
Second, it can easily be shown that the results are the same whether
one uses the ``bottom'' surface in Fig.~\ref{fig:gd} as a reference
and integrates up in $\lambda$, as done above, or chooses the ``top'' surface
as a reference and integrates down.
Third, the integration over the $\lam$ axis can
be carried out analytically in the basis that locally diagonalizes $B(\k)$,
as detailed in Appendix \ref{appen:int}.
Therefore only a 2D discrete integration over the \k\ plane is
needed, which is numerically efficient.
Lastly, in the single-band case all quantities such as
$W$, $B_{x}$ and $\Bbar_{x}$ obviously commute with each other, leaving
$G=G_1=\Tr\Big[B\O_{xy}^{(0)}\Big]$.
\footnote{A similar simplification
occurs in the multiband case if $B$ is globally diagonal (i.e., at
all $\k$), but this cannot normally be expected.}

In the following subsection, we discuss
the properties of the $\theta\GD$ formula
in the presence of $\cT$ symmetry, showing that if
a TR-symmetric gauge has
been chosen in the bulk BZ and assuming
that $B(\k)$ varies smoothly in the 2D \k\ plane,
both $\theta\BK$ and $\theta\GD$ must vanish.

%------------------------------------------------------------
\subsection{Time-reversal symmetry}
\label{sec:TR_tGD}
%------------------------------------------------------------

Consider the situation in which the system has $\cT$
symmetry and is topologically normal, and a gauge
respecting $\cT$ symmetry has been chosen
smoothly throughout  the bulk BZ
for the $2N$ occupied bands.
For such a system
we can construct $2N$ localized Wannier functions (WFs)
which fall into $N$ $\cT$-symmetric pairs,
\bea
&& \cT \ket{w_{n\mathbf{R},1}}=-\ket{w_{n\mathbf{R},2}}\;,\nn
&& \cT \ket{w_{n\mathbf{R},2}}=\ket{w_{n\mathbf{R},1}},
\label{eq:TR-wannier}
\eea
where $1\leq n\leq N$ is the index of a $\cT$-symmetric pair and
$\mathbf{R}$ denotes a real-space lattice vector. 
Typically $\ket{w_{n\mathbf{R},1}}$
and $\ket{w_{n\mathbf{R},2}}$ are chosen to diagonalize the
$S_z$ operator in their two-dimensional subspace,
%%%
\footnote{In general the spin quantization axis can
  be chosen to be different for different $\cT$-symmetric pairs.}
%%%
so that ``1" and ``2" can be interpreted roughly
as ``spin indices."
The Fourier transform of the
$\cT$-symmetric WF pairs leads to
a smooth gauge respecting $\cT$ symmetry in the bulk BZ,
\bea
&& \cT \ket{\psi_{n\k ,1}}\!=\!-\ket{ \psi_{n-\k ,2}} \;\nn
&& \cT \ket{\psi_{n\k ,2}}\!=\! \ket{\psi_{n-\k ,1}}
\label{eq:TRgauge-psi}
\eea
and
\bea
&& \cT \ket{u_{n\k ,1}}\!=\!-\ket{ u_{n-\k ,2}} \;\nn
&& \cT \ket{u_{n\k ,2}}\!=\! \ket{u_{n-\k ,1}}
\label{eq:TRgauge}
\eea
where the indices ``$1$" and ``$2$" are again the ``spin indices",
even if the directions of the spin expectation values can have some variations
with $\k$.
Note that the states in \eq{TRgauge-psi}
are of Bloch form, but in general are not
the eigenstates of the Hamiltonian.

Henceforth we shall say that a gauge that obeys
Eqs.~(\ref{eq:TRgauge-psi}-\ref{eq:TRgauge}) is a $\cT$-symmetric gauge.
However, in general a gauge obeying these equations
is not necessarily periodic.
For example, there may be a gauge discontinuity located at
some boundary plane in the 3D BZ. When the \Z2\
index of the system is even, such
a gauge discontinuity can typically be removed by
smoothening the gauge
without breaking the $\cT$ symmetry.
When the \Z2\ index is odd, however, the gauge discontinuity can never be eliminated
without breaking the $\cT$ symmetry in the gauge.
For, if it could, one could again construct $\cT$-respecting WFs,
which is known to
be impossible for \Z2-odd insulators.

If the gauge in the bulk BZ satisfies \eq{TRgauge},
it follows that the Berry curvatures
and Berry connections obey
\bea
&& A_a(\k) \! =\!\sigma_y\,\Big(A_a(-\k)\Big)^{T}\,\sigma_y ,\;\nn
&& \O_{ab}(\k)\!=\!-\sigma_y\,\Big(\O_{ab}(-\k)\Big)^{T}\,\sigma_y \;,
\label{eq:berry-trgauge}
\eea
where $a$ and $b$ run over the reciprocal-lattice directions.
All the quantities in \eq{berry-trgauge} are $2N\times 2N$ matrices.
In particular, $\sigma_y$ denotes the outer product between the $2\times 2$ Pauli matrix
$\tau_y$ and the $N\times N$ identity matrix, and
the superscript ``$T$" refers to matrix transpose for the $2N\times 2N$ matrices.
Since the Berry curvature is odd in $\k$,
while the Berry connections behave
as even functions of $\k$,
it is easy to show that both $\Tr\Big[\,A_a(\k)\,\O_{bc}(\k)\,\Big]$
and $\Tr\Big[\,i A_a(\k)\,[A_b(\k), A_c(\k) ]\,\Big]$
are canceled by their time-reversal partners at $-\k$. Therefore, the bulk
integral $\theta\BK$ in Eq.~(\ref{eq:theta-b}) vanishes if a smooth
$\cT$-respecting gauge is constructed in the bulk BZ.

In particular, at the boundary plane where the
gauge discontinuity is located,
the wavefunctions at the bottom and top planes (say, $k_z=\pm\pi$)
are connected
via $\cT \ket{u^{(0)}_{n\k,1}}\!=\!-\ket{ u^{(1)}_{n-\k ,2}}$ and
$\cT \ket{u^{(0)}_{n\k,2}}\!=\!\ket{ u^{(1)}_{n-\k ,1}}$,
where $\k$  is now understood to be a wavevector in the 2D plane.
With such a $\cT$-respecting gauge choice,
the $B$ matrix, the Berry connections, and the Berry curvature
satisfy the following relationships:
\bea
&& B(\k)=\sigma_y B(-\k)^{T}\sigma_y \;,\;
\label{eq:trsym1}\\
&& \Ab_x(\k)=\sigma_y\,\Big(\At_x(-\k)\Big)^{T}\,\sigma_y \;,\;
\label{eq:trsym2}\\
&& \Ab_y(\k)=\sigma_y\,\Big(\At_y(-\k)\Big)^{T}\,\sigma_y \;,\;
\label{eq:trsym3}\\
&& \O^{(0)}_{xy}(\k)=
-\sigma_y\,\Big(\O^{(1)}_{xy}(-\k)\Big)^{T}\,\sigma_y\; .
\label{eq:trsym4}
\eea
Again, superscripts $``(0)"$ and $``(1)"$
refer to the quantities evaluated
at $\lambda\!=\!0$ and $\lambda\!=\!1$ respectively.
We now show that if Eqs.~(\ref{eq:trsym1})-(\ref{eq:trsym4})
are satisfied, and if all the quantities involved in the
Eqs.~(\ref{eq:Gcurly})-(\ref{eq:Gfull}) vary smoothly in the 2D plane,
then $\theta\GD$ must vanish.

First of all, it is straightforward to show that the first term  in
\eq{Gcurly} vanishes due to $\cT$ symmetry.
As the gauge-covariant Berry curvature
on the top plane ($\lam=1$) is connected the one on the bottom plane
($\lam=0$) via $\Ot^{(1)}_{xy}\!=\!U\dag\,\Ot^{(0)}_{xy}\,U$,
and $U\!=\!e^{-iB}$ commutes with $B$, it follows that
$\Tr\Big[B(\k)\Ot^{(1)}_{xy}(\k)\Big]\!=\!\Tr\Big[B(\k)\Ot^{(0)}_{xy}(\k)\Big]$.
On the other hand, from \eq{trsym1} and \eq{trsym4} we know that
$\Tr\Big[B(\k)\Ot^{(0)}_{xy}(\k)\Big]\!
=\!-\Tr\Big[B(-\k)\Ot^{(1)}_{xy}(-\k)\Big]$,
which leads to an exact cancellation for the first term.
%we know $B$ transforms as an even function of $\k$,
%while $\Ot_{xy}$ is odd in $\k$,  the first term $G_1(\k)$ (\eq{G1final})
%is thus exactly cancelled by $G_1(-\k)$.

The second term in \eq{Gcurly} is trickier.
First,  from the representation-invariance of the trace
and the fact that $W=e^{-i\lam B}$ commutes with $B$,
we know that $\Tr\Big[i\,B \,[\calA_x,\calA_y]\Big]\!
=\Tr\Big[i\,B\,[A_x^{(\lam)}, A_y^{(\lam)}]\Big]$. Then we claim
that the Berry connection matrix at $(\k, \lam)$
is connected to the one at $(-\k, 1-\lam)$ via a
$\cT$ transformation
\beq
A_{a}^{(\lam)}(\k)\!=
\!\sigma_y\,\Big(A_{a}^{(1-\lam)}(-\k)\Big)^{T}\,\sigma_y\; ,
\label{eq:Atrsym}
\eeq
where $A_{a}^{(\lam)}\equiv A_{a}(\lam)$ with $a=\{x, y\}$
as defined in Eq.~(\ref{eq:Ax})-(\ref{eq:Ay}).
\eq{Atrsym} will be proved properly in Appendix \ref{appen:trsym},
but if one considers
$\lam$ as the third wavevector component,
\eq{Atrsym} is indeed very intuitive.
Combing \eq{Atrsym} and \eq{Alxy}, it follows that
\beq
%\Tr\Big[i\,B(\k)\,[\calA_x^{(\lam)}(\k),\calA_y^{(\lam)}(\k)]\Big]\;\nn
%=\,-\Tr\Big[i\,B(-\k)\,
%[\calA_x^{(1-\lam)}(-\k)\,,\,\calA_y^{(1-\lam)}(-\k)]\Big]
\eta(\k,\lambda)=-\eta(-\k, 1-\lam)\;,
\eeq
where
\bea
\eta(\k,\lambda)&=&\Tr\Big[\,i\,B(\k)\,[\calA_x^{(\lam)}(\k),
\calA_y^{(\lam)}(\k)]\,\Big]\;\nn
&=&\Tr\Big[\,i\,B(\k)\,[A_x^{(\lam)}(\k),
A_y^{(\lam)}(\k)]\,\Big]
\eea
is exactly the second term in \eq{Gcurly}.
Therefore, that term
also vanishes due to the cancellation
between the integrands at $(\k, \lam)$ and $(-\k, 1-\lam)$.
It thus follows that $\theta\GD$ has to vanish
for a $\cT$-respecting gauge choice.

%------------------------------------------------------------
\section{Vortex-loop contribution}
\label{sec:vl}
%------------------------------------------------------------

In the previous section, we derived a formula for the gauge
discontinuity contribution $\theta\GD$, as expressed in
Eq.~(\ref{eq:tGD}) and Eqs.~(\ref{eq:Gcurly})-(\ref{eq:Gfull}).
We also demonstrated that for a system with
$\cT$ symmetry, if a $\cT$-respecting gauge
is constructed in the bulk BZ, and if the branch
choice is made in such a way that $B(\k)$ varies
smoothly over the entire 2D \k\ plane, then both $\theta\BK$ and
$\theta\GD$ must vanish.

However, it is well known that $\theta=\pi$ for \Z2\ TIs,
so one may wonder where the quantized $\theta$ can come from?
The answer is that, in the \Z2-odd\ case, it is topologically
impossible to insist on a branch choice such that $B$ remains
smooth throughout the $(k_x, k_y)$ plane.  In other words, the
2D \k\ plane has to be subdivided such that one or more of the
eigenvalues of $B$ change by an integer multiple of $2\pi$ when
crossing from one subregion to another.  We denote the boundaries
of such 2D subregions as ``vortex loops.''  It turns out that
the vortex-loop contribution is exactly $\pi$ for a \Z2\ TI.

In this section, we introduce such vortex loops and discuss their
contribution to the $\theta$ coupling.  We first propose a formal
definition of a vortex loop in Sec.~\ref{sec:vldef}, and then
derive a formula for the vortex-loop contribution $\theta\VL$
in Sec.~\ref{sec:vl-derive}.  This formula turns out to be rather
simple, involving two Berry phases that are accumulated as one
traverses the vortex loop, one associated with the electronic
Bloch-like functions and the other with the eigenvectors
of $B(\k)$.  In Sec.~\ref{sec:vl-discuss} we discuss several
properties of our formula for $\theta\VL$. In particular, we show
that in systems with $\cT$ symmetry, and for which the gauge also
respects $\cT$ symmetry, $\theta\VL$ must be either 0 or $\pi$,
corresponding to the \Z2\ classification of 3D $\cT$-invariant
insulators.

\subsection{What is a vortex loop}
\label{sec:vldef}
%------------------------------------------------------------------------

In Sec.~\ref{sec:statement} we suggested that the complete formula
for $\theta$ should include three kinds of contributions, as expressed
by \eq{thetatot}.  Here we review the philosophy of the calculation,
explaining why the third vortex-loop contribution $\theta\VL$ may be
needed.

First, we choose a smooth gauge in the 3D bulk BZ, but the periodicity condition
in the $k_z$ direction is relaxed. Hence some gauge discontinuity is introduced
at a 2D boundary plane normal to $k_z$.
The 3D bulk integral of \eq{theta-a} (excluding the boundary plane)
is the $\theta\BK$ term in \eq{thetatot}.

\begin{figure}
\centering
\includegraphics[width=6cm]{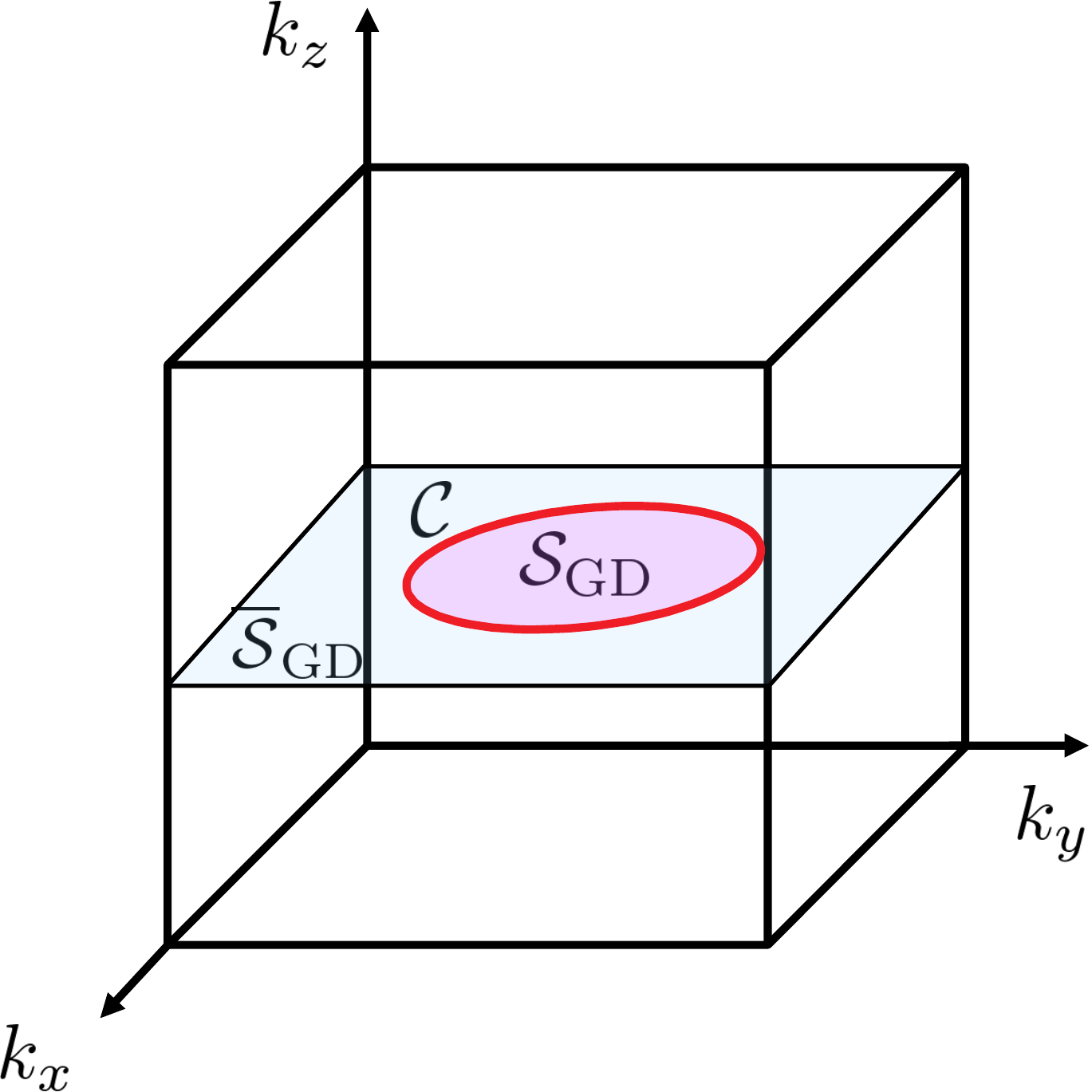}
\caption{\label{fig:2patch} Schematic illustration of a 3D
BZ containing a 2D plane of gauge discontinuity
that is divided into two patches $\cS_{\textrm{GD}}$ and
$\overline{\cS}_{\textrm{GD}}$ by a vortex loop
(red line).}
\end{figure}

Next, we identify the 2D boundary as $\cS$ .
Let us define $\cS$ as a directed
area with surface normal $\hat{\mathbf{n}}$. In order to compute the integral
over the 2D plane $\cS$, $\hat{\mathbf{n}}$ is chosen
in such a way that $\hat{\mathbf{x}}$-$\hat{\mathbf{y}}$-$\hat{\mathbf{n}}$
form a right-handed coordinate triad.  The gauge discontinuity
in the $\hat{\mathbf{n}}$ direction is given by a unitary matrix
$U(\k)=e^{-iB(\k)}$ which varies smoothly with $\k$ lying in the 2D plane.
Since the Hermitian matrix $B(\k)=i\ln U(\k)$ is involved in the formula
for $\theta\GD$ (\eq{Gfull}), a branch choice for $B$ has to be made.
If possible we make a branch choice so that $B(\k)$ is
smooth and continuous over the entire \k\ plane, but this may not
always be possible or desirable.  In that case $\cS$
is divided into subregions within
each of which $B(\k)$ is smooth and continuous. For example,
Fig.~\ref{fig:2patch} shows $\cS$ divided into two subregions
$\cS_{\GD}$ and $\overline{\cS}_{\GD}$ separated by a boundary loop
$\mathcal{C}$, which we refer to as a ``vortex loop."
The 2D contribution $\theta\GD$ is then computed by integrating over
all subregions of $\cS$ using
Eq.~(\ref{eq:Gcurly})-(\ref{eq:Gfull}) of Sec.~\ref{sec:surf}.

Since the $B$ and $U$ matrices have the same eigenvectors,
the eigenvalues of $B$ may exhibit abrupt $2\pi$ jumps as
they vary from one subregion to another
(from $\mathcal{S}_{\rm{GD}}$ to $\mathcal{\overline{S}_{\rm{GD}}}$
in Fig.~\ref{fig:2patch}),
even though $U$ remains smooth throughout the $(k_x, k_y)$ plane.
The behavior of $B$ is thus singular when crossing the vortex loops.
The vortex-loop contributions cannot be computed
from the formula for $\theta\GD$;
a new formula is needed to account for them.

In more general cases, a 3D BZ may be divided into multiple
subvolumes, and these
subvolumes can meet on multiple 2D surface patches with gauge discontinuities.
These surface patches may further meet at one or more 1D lines or
curves, which may behave as vortex loops.
For such cases the definition of a vortex loop would need to be
generalized, since the $U(\k)$ matrices obtained by approaching
the meeting line from different patches are in general no longer
consistent, and may not even commute with one another.
We leave this more complicated situation
to a future study.

The presence or absence of vortex loops clearly depends on how
the branch choice of $B$ is made in the 2D \k\ plane. We normally
try to make this choice so as to avoid vortices.  If the system does not
have $\cT$ symmetry (and assuming vanishing Chern
numbers), then it is usually straightforward to do this,
since the eigenvalues of $B$ typically remain non-degenerate
throughout the 2D \k\ plane (degeneracies in a general
Hermitian matrix are of codimension three, and so do not occur
without special tuning in a 2D \k\ plane).

However, when the system is topologically nontrivial,
this may become impossible; a topological
obstruction may force the existence of at least one vortex loop.
In particular, if $\cT$ symmetry is present,
there must be a degeneracy between two different eigenvalues of $B$ at
the four time-reversal invariant momenta (TRIM) in the
2D \k\ plane.
\footnote{Let us consider the gauge-discontinuity plane as
an isolated 2D BZ without worrying about its $k_z$ value.}
As a result, the topological properties of the bulk Hamiltonian
become closely related to the number
of vortex loops. In the same vein as the \Z2\ classification
based on the number of surface Dirac cones,\cite{fu-prl07}
when there is an odd number of vortex loops,
the system is \Z2-odd, corresponding
to a $\cT$-respecting topological insulator.
Otherwise when the number of vortex loops is even,
the system is topologically trivial. In the topologically
nontrivial case,
it is impossible to insist on the smoothness of all of the
eigenvalues of $B$ throughout the 2D $\k$ plane.
In principle the last vortex loop can be made infinitesimally small
by shrinking it around one
of the TRIM, but
the symmetry-protected degeneracy at the TRIM prevents it from
being removed completely.
Therefore, we must consider the contribution from vortex loops
in such topologically nontrivial phases.

On the other hand,
vortex loops may be present even in topologically trivial cases
unless one makes a proper branch choice to remove them.
In realistic calculations, for example, one usually adopts some
default branch choice for the eigenvalues of $B$ (e.g., from -$\pi$ to $\pi$),
which is not necessarily  the one that makes $B$ globally smooth.
In such cases one has to consider both $\theta\GD$ and $\theta\VL$.
In this regard,
it would be useful to have a formula for the vortex loop contribution,
so that one can evaluate the gauge-discontinuity contribution to $\theta$
for an arbitrary branch choice.

In the remainder of this section we will derive a formula for $\theta\VL$
and discuss properties of the formula.  We will also show that in the
presence of $\cT$ symmetry in both the Hamiltonian and the
gauge, the vortex-loop contribution alone determines whether the system
is \Z2-odd\ ($\theta\VL=\pm\pi$) or \Z2-even\ ($\theta\VL=0$).

%------------------------------------------------------------
\subsection{The formula for $\theta\VL$}
\label{sec:vl-derive}
%------------------------------------------------------------

Let us first consider the topologically trivial case
in which we can always find a proper branch choice such that $B$ remains
smooth throughout the 2D plane. Assuming this has been done, now shift
the $n$th eigenvalue of
$B$ by $2\pi\nu(n)$ within subregion ${\cS}_{\GD}$,
thus creating
a vortex loop $\mathcal{C}$ whose interior is 
${\cS}_{\GD}$ 
as shown in Fig.~\ref{fig:2patch}.
The above operation is equivalent to making
a different branch choice. However, a physical quantity should
be  independent of the branch choice,
so $\theta$ should remain invariant after such an operation.
Letting $\theta_{\textrm{shift}}$ be the change in $\theta\GD$
arising from this redefinition of $B$ in the interior region
${\cS}_{\GD}$, 
it follows that we must have
\begin{equation}
\theta\VL=-\theta_{\textrm{shift}}  \;.
\label{eq:shift}
\end{equation}

We begin by considering a simple case in
which only one of the eigenvalues of $B$
is shifted by $2\pi$ within ${\cS}_{\GD}$.
We make the decomposition  $B=B_0+\Delta B$, where
$B_0$ is the original smooth part and $\Delta B$ is the
change arising from the $2\pi$ shift.
We then choose to connect the states at the
bottom and top planes in two steps. In the first step,
\beq
\ket{\psi^{(\lam)}_m}=\sum_{n=1}^{N}\ket{\psi^{(0)}_n}\,(e^{-i\lam \D B})_{nm}, \;\;\;
\lam\in [0,1).
\label{eq:step1}
\eeq
In the second step,
\beq
\ket{\psi^{(\lam)}_m}=\sum_{n=1}^{N}\ket{\psi^{(1)}_n}\,(e^{-i(\lam-1) B_0})_{nm},\;\;\;
\lam\in [1,2].
\label{eq:step2}
\eeq
Note that the states at the top plane are now denoted as
$\ket{\psi^{(2)}_m}$ instead of $\ket{\psi^{(1)}_m}$.
In the second step, $B_0$ is smooth over the entire 2D BZ;
one can define $\lambda'=\lambda-1$ with $\lambda'\in [0, 1]$,
and the formula for $\theta\GD$ derived in Sec.~\ref{sec:surf}
applies.  Thus, $\theta_\textrm{shift}=-\theta\VL$ is just
the contribution to $\theta\GD$ coming from the gauge twist of
\eq{step1} in the loop interior 
${\cS}_{\GD}$.

We assume without
loss of generality that the first
eigenvalue of $B$ (denoted by $b_1$) jumps by $2\pi$ in
the subregion ${\cS}_{\GD}$.
Then $\Delta B$ can be written as
\bea
\Delta B =
\begin{cases}
  V\,\Delta_1 \,V^\dagger & \text{if }\k\in{\cS}_{\GD} \\
  0                     & \text{otherwise}
\end{cases}
\label{eq:dB}
\eea
where $\Delta_1$ is an $N\times N$ matrix
($N$ is the number of occupied bands), with $(\Delta_1)_{11}=2\pi$
and all the remaining matrix elements vanishing.
Here $V=(v_1, v_2,..., v_N)$
is the unitary matrix whose $n$'th column $v_n$ is the $n$'th eigenvector
of $B_0$.  Plugging this expression for $\D B$ into
the expression for $G$ in Eq.~(\ref{eq:Gfull}), one obtains
a formula for $\theta_{\textrm{shift}}$, and $\theta\VL$ is
simply the opposite of $\theta_{\textrm{shift}}$.
After some considerable algebra, which we defer to
Appendix~\ref{appen:vl}, it turns out that many terms cancel,
and one obtains the surprisingly simple formula
\beq
\theta\VL=-\theta_{\textrm{shift}}=\Big[\,\phi_1(\mathcal{C})+
\xi_1({\mathcal{C}})\,\Big]/2 \; .
\label{eq:vl1}
\eeq
Here $\phi_1$ and $\xi_1$ are two different, but related, Berry
phases that need to be computed around loop $\mathcal{C}$ (taking
the positive
sense of circulation with respect to the unit normal to ${\cS}_{\GD}$).
The second term $\xi_1({\mathcal{C}})$ is easier to describe; it
is just the Berry phase of the $N$-component vector $v_1$
(the first column of $V$) as it is taken around the loop $\mathcal{C}$.
To understand the first term, note that the elements of $V$ can be
used to build the linear combinations
\beq
\ket{\bar{\psi}^{(0)}_{n\k}}=\sum_{m=1}^{N} \ket{\psi^{(0)}_{m\k}}\,V_{mn}
\label{eq:psibar}
\eeq
out of the Bloch functions $\ket{\psi^{(0)}_{m\k}}$ at the bottom
plane ($\lambda\!=\!0$), such that
\beq
\ket{\bar{\psi}^{(0)}_{1\k}}=\sum_{m=1}^{N} v_{1,m} \, \ket{\psi^{(0)}_{m\k}}
\eeq
is precisely the state whose phase is shifted by $2\pi$, while
the other $N-1$ states are unaffected by $\Delta B$.  Then the first
term $\phi_1(\mathcal{C})$ in \eq{vl1} is just the Berry phase of
$\ket{\bar{\psi}^{(0)}_{1\k}}$ as it is carried around the loop
$\mathcal{C}$.  The gauge-invariance and other properties
of this formula will be discussed further in the next subsection.

In the most general case, there may be multiple vortex
loops $\{\mathcal{C}_i, i=1,... L\}$ in the 2D \k\ plane, and inside
the $i$th vortex loop the $n$th eigenvalue of $B$
may be shifted by $2\pi\nu_{n}(i)$ with $\nu_{n}(i)$ being an integer.
Then Eq.~(\ref{eq:vl1}) can be generalized in
a straightforward manner to
\bea
\theta\VL=\sum_{i}\sum_{n} \, \frac{\phi_{n}(\mathcal{C}_i) +
\xi_{n}(\mathcal{C}_i)}{2}\;\nu_n(i)
\label{eq:vl_final}
\eea
where $\phi_{n}(\mathcal{C}_i)$ and $\xi_{n}(\mathcal{C}_i)$
are the Berry phases around the loop $\mathcal{C}_i$
of the $n$th Bloch-like state $\ket{\bar{\psi}^{(0)}_{n\k}}$ (\eq{psibar})
and the $n$th eigenvector of $B$ respectively.
\eq{vl1}, together with its generalized form \eq{vl_final},
is the other central result of this paper.

%--------------------------------------------------------------------------------------
\subsection{Discussion}
\label{sec:vl-discuss}
%--------------------------------------------------------------------------------------

We discuss the properties of
\eq{vl_final} in this subsection.
We first show that  \eq{vl_final} is indeed
gauge invariant modulo $2\pi$, which is consistent with
the $2\pi$ ambiguity of $\theta$. Secondly we prove that
\eq{vl_final} remains unchanged by interchanging the
two steps corresponding to Eqs.~(\ref{eq:step1}) and (\ref{eq:step2}).
Lastly we discuss the case of $\cT$ symmetry and conclude that
as long as a gauge respecting $\cT$ symmetry is used,
$\theta\VL=\pm\pi$ or 0 depending on whether the system is
\Z2-odd\ or \Z2-even, respectively.

%-----------------------------------------------------------------
\subsubsection{Gauge invariance}
%-----------------------------------------------------------------

\eq{vl_final} is rather unexpected, as it involves the
average of two Berry phases in a manner that, to our knowledge, has
not been encountered before.  Nevertheless, it is easy to confirm
that it obeys one important property, namely, that it is well-defined
modulo $2\pi$, as required for any plausible formula for $\theta$.
To prove this, we  first note
that the only gauge freedom in \eq{vl_final} is
a $U(1)$ gauge transformation acting on $v_n$, i.e.,
$v_n\to v_n e^{i\beta}$ ($\k$-dependence is implicit).
On the other hand,
since $\ket{\bar{u}^{(0)}_n}=\sum_{m=1}^{N}\ket{u^{(0)}_m}\,v_{n,m}$,
the same gauge transformation must also be applied to
$\ket{\bar{u}^{(0)}_n}$, i.e.,
$\ket{\bar{u}^{(0)}_n}\to\ket{\bar{u}^{(0)}_n}e^{i\beta}$.
As a result, if the gauge transformation has a
non-zero winding number $J$, such that
$\xi_n$ is changed by $2\pi J$, then
$\phi_n$ must change by $2\pi J$ as well.
It follows that Eq.~(\ref{eq:vl_final})
is gauge invariant modulo $2\pi$.

%%-----------------------------------------------------------
\subsubsection{Order of the two steps}
%%-----------------------------------------------------------

In Sec.\ref{sec:vl-derive} we decomposed $B$ into two parts,
$B=B_0+\D B$, where $B_0$ is the smooth part and $\D B$ is the
contribution from the $2\pi$ shift (equal and opposite to the
vortex-loop contribution).  Then, in Eqs.~(\ref{eq:step1}) and
(\ref{eq:step2}), $B$ was treated in two steps in the fictitious
$\lambda$ space.  The first step ($0\!<\!\lambda\!<\!1$) dealt
with  $\D B$, while the second step ($1\!<\!\lambda\!<\!2$)
treated the smooth part $B_0$.  Here we would like to show
that Eqs.~(\ref{eq:vl1}) and (\ref{eq:vl_final}) remain correct
regardless of the order of the $\lambda$ integrations.

If the order is reversed,
it is straightforward to show that Eq.~(\ref{eq:vl1})
remains unchanged, but the first term $\phi_1$ is
interpreted as the Berry phase of $\ket{\bar{u}^{(1)}_1}$,
where $\ket{\bar{u}^{(1)}_1}\!=\!\sum_{n}\ket{u^{(1)}_n}\,v_{1,n}\!=\!
\sum_{n,m}\ket{u^{(0)}_m}\,(e^{-iB_0})_{mn}\,v_{1,n}$.
The Berry phases of $\ket{\bar{u}^{(1)}_1}$ and
$\ket{\bar{u}^{(0)}_1}$ around the vortex loop $\mathcal{C}$
are exactly the same,
because $\ket{\bar{u}^{(1)}_1}\!=\!\sum_{m,n,l}
\ket{\bar{u}^{(0)}_l}\,v_{l,m}^*\,(e^{-iB_0})_{mn}\,v_{1,n}\!=\!
\ket{\bar{u}^{(0)}_1}\,e^{-ib_1}$, where $b_1$ is the first eigenvalue
of $B_0$. Since $b_1$ is smooth and single-valued everywhere in the
2D $\k$-plane,
the Berry phase would not change under such a single-band transformation.
Therefore, \eq{vl1} and \eq{vl_final} remain valid even if the
order of \eq{step1} and \eq{step2} is reversed.

\subsubsection{Time-reversal symmetry}
\label{sec:tr}

We  proceed to prove that $\theta\VL$ must be either
$\pm\pi$ or $0$ for $\cT$-invariant systems
when the gauge in the bulk BZ
is chosen to respect $\cT$ symmetry. Again, let us consider the simple
case that there is only one vortex loop $\mathcal{C}$ in
the 2D \k\ plane, and that only the first eigenvalue of $B$
is shifted by $2\pi$ inside the vortex loop. Suppose that a smooth gauge
respecting $\cT$ symmetry has been constructed in the bulk BZ,
so that both the bulk integral $\theta\BK$ and the surface integral
$\theta\GD$ vanish as discussed in Sec.~\ref{sec:TR_tGD}.
Due to the $\cT$-respecting gauge of Eq.~(\ref{eq:TRgauge}),
the $B$ matrix must satisfy Eq.~(\ref{eq:trsym1}),
with two eigenvalues being degenerate at each of the four TRIM, i.e.,
$(0, 0)$, $(0, \pi)$, $(\pi, 0)$ and $(\pi, \pi)$. As a result,
the vortex loop $\mathcal{C}$ has to be a ``$\cT$-symmetric" loop centered
at one of the TRIM, which means that
for any $\k$ on the loop $\mathcal{C}$, $-\k$
must also lie on the loop.
Then it is well known that the Berry phase around such
a $\cT$-symmetric loop enclosing a degeneracy point is $\pm\pi$,
as has been demonstrated in the surface states of TIs and
in $\cT$-invariant systems with giant Rashba spin-orbit
splitting.\cite{shen-prb04,tokura-science13} It follows
that $\xi_1\!=\!\pm\pi$ in Eq.~(\ref{eq:vl1})

It can be further shown that $\phi_1$ in Eq.~(\ref{eq:vl1})
is exactly the same as $\xi_1$ as a result of
the $\cT$ symmetry. Let us first make a branch
choice such that the vortex loop
is negligibly small,
then the Berry connection of $\ket{\bar{u}_1^{(0)}}$
can be expressed as
\bea
\bar{A}_{a,11}^{(0)}
&=&i\,\langle{\bar{u}_1^{(0)}}\vert{\partial_a}
{\bar{u}_{1}^{(0)}}\rangle \;\nn
&=&i \sum_{m,n=1}^{N}\!V_{m1}^{*}\langle{u_{m}^{(0)}}
\vert{\partial_a u_{n}^{(0)}}\rangle\,V_{n 1}+
i \sum_{n=1}^{N} \,V_{n1}^{*}\parx V_{n1} \;\nn
&=&( V^{\dag}\!A_{a}^{(0)}\,V )_{11}+ C_{a,11}
\eea
where $N$ is the number of occupied
bands, $A_{a}^{(0)}$ is the Berry-connection matrix in
the bottom-plane gauge with $a\!=\!\{x, y\}$, and
\beq
C_{a,mn}=i\sum_{l=1}^{N}V_{lm}^{*}\partial_a V_{ln}\; ,
\label{eq:Cxy}
\eeq
may be  interpreted as the ``Berry connection" in the gauge space.
As the vortex loop is chosen to be vanishingly small,
the variation of $\ket{u^{(0)}_1}$ within
the vortex loop is negligible.
Therefore $A_{a}^{(0)}\!=\!0$,
which means $\bar{A}_{a,11}^{(0)}$ comes only from
the gauge twist, i.e.,
$\bar{A}_{a,11}^{(0)}\!=\!C_{a,11}$. It follows that
$\xi_1\!=\!\phi_1\!=\!\pm\pi$ for such a special branch choice,
and $\theta\VL\!=\!\pm\pi$ according to Eq.~(\ref{eq:vl1}).

Now suppose the loop is enlarged while preserving
$\cT$ symmetry in the shape of the loop.  We showed
at the end of Sec.~\ref{sec:tr} that contributions to
$\theta\GD$ coming from $\k$ and $-\k$ always cancel when there
is a $\cT$-respecting gauge in the bulk, so $\theta\GD$ continues
to vanish as the loop is enlarged.  By the argument given around
\eq{shift}, this means $\theta_\textrm{shift}$, and therefore
$\theta\VL$, cannot change as the loop is enlarged, even if
the variation of $\ket{u_{1}^{(0)}}$
is no longer negligible.
In other words,
given a $\cT$-respecting gauge in the bulk
BZ, $\theta\VL$ must be quantized as $\pm\pi$ in the \Z2-odd\ case regardless
of the size of the vortex loop.

We can generalize the discussion to
a more general case with multiple vortex loops. Obviously
when there is an odd number of vortex loops,
 $\theta$ is still quantized as $\pm\pi$ (modulo $2\pi$).
If there is an even number of vortex loops, they can either enclose an even
number of TRIM or fall into $\cT$ partners without enclosing any
TRIM, and $\theta$ has to vanish (modulo $2\pi$) in either case.

%%%%%%%%%%%%%%%%%%%%%%%%%%%%%%%%%%%%%%%%%%%%%%%%%%%%%%%%%%%%%%%%%%
\section{Applications}
\label{sec:apply}

In this section, we apply our method to the Fu-Kane-Mele (FKM)
model,\cite{fu-prl07} which is a 4-band tight-binding
model of $s$ electrons on the diamond lattice.
The model Hamiltonian is
\beq
H=\sum_{\langle i,j\rangle} t_{ij} c_i\dag c_j+
i8\lam_{\textrm{SO}}\sum_{\langle\langle i,j\rangle\rangle}
c_i\dag\,\mathbf{s}\cdot(\mathbf{d}_{ij}^1\times\mathbf{d}_{ij}^2)\,c_j\;,
\eeq
where $t_{ij}$ is the first-neighbor spin-independent hopping and
$\lam_{\textrm{SO}}$ is the strength of the second-neighbor spin-dependent
hopping generated by spin-orbit coupling (SOC);
$\mathbf{d}_{ij}^1$ and $\mathbf{d}_{ij}^2$
are the two first-neighbor bond vectors connecting
the two second-neighbor sites $i$ and $j$; and $\mathbf{s}\!=\!(s_x, s_y, s_z)$ are
Pauli matrices representing the electronic spin.
Hereafter we only consider the case of half filling, i.e., two occupied bands.
Setting $t_{ij}\!=\!t_0\!=\!1$ and $\lam_{\textrm{SO}}\!=\!0.125$,
it is easy to check that
the system is a semimetal with gap closures at the
three equivalent $X$ points in the BZ
when the diamond-lattice symmetry is preserved.
An energy gap can be opened up if an appropriate
symmetry-lowering perturbation is added. For example,
when the first-neighbor bond along the
[111] direction is distorted,
the system can be either a trivial insulator or a
topological insulator depending on the strength of
the bond distortion.

In order to validate our method, we need to consider the general
case without $\cT$ symmetry. Following  Ref.~\onlinecite{essin-prl09},
we modify the system by applying a staggered Zeeman field 
with amplitude $h$, direction $\hat{\mathbf{h}}$ along [111],
and opposite signs on the  $A$ and $B$ sublattices.
Moreover, the [111] first-neighbor bond is distorted by
changing the corresponding hopping amplitude from
$t_0$ to $3t_0+\delta$. We work in polar coordinates
in the $(\delta,h)$ parameter space, i.e.,
$\delta\!=\!m \cos\beta$ and $h\!=\!m \sin\beta$. The Hamiltonian then becomes
\bea
H(\beta)&=& \!\! \sum_{\langle i,j\rangle={[1 1 1]}}
\!\!\! (3t_0+m \cos\beta)\, c_i\dag c_j  + \!\!\!\!
\sum_{\langle i,j\rangle\neq{[1 1 1]}} \!\!\! t_0\,c_i\dag c_j \nn
& +&i8\lam_{\textrm{SO}}\sum_{\langle\langle i,j\rangle\rangle}
c_i\dag\,\mathbf{s}\cdot(\mathbf{d}_{ij}^1\times\mathbf{d}_{ij}^2)\,c_j \nn
&+& m \sin\beta \sum_{i} c_i\dag \,
  \hat{\mathbf{h}}\cdot\mathbf{s}\;\tau_z \, c_i
\eea
where $\tau_z$ is the Pauli matrix defined in the space of the two sublattices.
When $\beta\!=\!0$ and $\pi$, the Zeeman field vanishes so that $\cT$ symmetry
is restored, but the
topological index reverses between these two cases.
As $\beta$ increases from $0$ to $\pi$, the system varies smoothly from
a trivial to a topological insulator along a $\cT$-breaking path
without closing the bulk energy gap.

\begin{figure}
\includegraphics[width=3.4in]{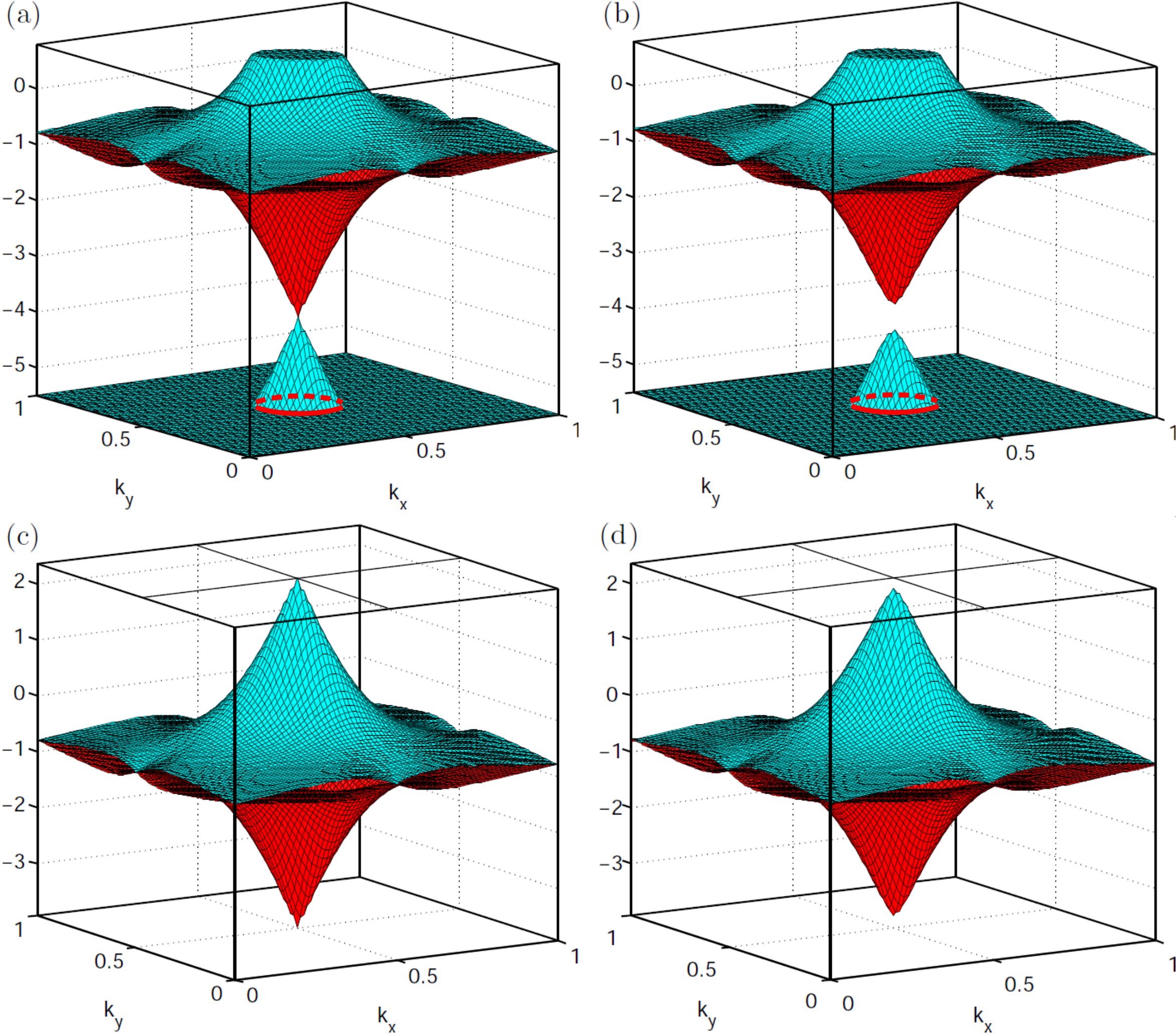}
\caption{3D plots of the two eigenvalues (colored red and cyan) of
$B(k_x, k_y)\!=\!i\,\ln U(k_x, k_y)$  for the Fu-Kane-Mele model
at half filling: (a) when the system is a TI, i.e., $\beta\!=\!\pi$,
with the branch choice taken as $(-7\pi/4, \pi/4]$;
(b) when $\beta\!=\!0.95\pi$, with the branch choice
$(-7\pi/4, \pi/4]$; (c) when $\beta\!=\!\pi$, with the
branch choice $(-5\pi/4, 3\pi/4]$; and
(d) when $\beta\!=\!0.95\pi$, with the
branch choice $(-5\pi/4, 3\pi/4]$.
The wavevectors $k_x$ and $k_y$ are
reduced such that $k_j\in[0,1]$ generates the 2D BZ.}
\label{fig:b1b2}
\end{figure}

Setting $t_0\!=\!1$, $\lam_{\textrm{SO}}\!=\!0.125$,
and $m\!=\!0.5$, we first study the behavior  of the $B$ matrix
of \eq{U} in the $(k_x , k_y)$ plane with the branch choice
$(-7\pi/4,\pi/4]$ for the eigenvalues of $B(\k)$.
As shown in Fig.~\ref{fig:b1b2}(a), when the system is
in the \Z2-odd\ phase ($\beta\!=\!\pi$)
there is a single vortex loop
surrounding the TRIM at $(\pi, \pi)$. Within the loop,
one of the eigenvalues of $B$ (shown in cyan) is shifted by $2\pi$, while
the other eigenvalue remains continuous. Moreover,
as a result of $\cT$ symmetry, the two eigenvalues of $B$
are degenerate at each TRIM, leading to quantized Berry
phases as discussed in Sec.~\ref{sec:vl-discuss}.
Fig.~\ref{fig:b1b2}(b) shows what happens if
$\cT$ symmetry is broken by setting $\beta\!=\!0.95\pi$.
Even though the vortex loop is still present for this value of
$\beta$, the two eigenvalues of $B$ are no longer degenerate
at the TRIM.

As discussed in Sec.~\ref{sec:vldef}, for the \Z2-odd\ case
a vortex loop has to be present
regardless of the branch choice.
The best one can do is to compress
the vortex loop to one of the TRIM in the 2D plane.
This is illustrated in Fig.~\ref{fig:b1b2}(c), where
the system of Fig.~\ref{fig:b1b2}(a) is reanalyzed using a
branch choice of
$(-5\pi/4,3\pi/4]$.
Now
the vortex loop is compressed to the point
$(\pi, \pi)$ in the $(k_x , k_y)$
plane. On the other hand, using the
same branch choice, the vortex loop can be completely removed
when $\beta\!=\!0.95\pi$,
as shown in Fig.~\ref{fig:b1b2}(d).

Using the methods developed in Secs.~\ref{sec:surf} and \ref{sec:vl},
we have calculated the total axion response $\theta$ along
the path from $\beta\!=\!0$ to $\pi$ by taking the sum of
$\theta\BK$, $\theta\GD$ and $\theta\VL$.
We first explain the
procedures for these calculations before discussing any specific results.
The parallel-transport technique,\cite{MLWF-1} which is detailed
in Appendix \ref{appen:pt}, is heavily used in the gauge construction.
As discussed earlier,
the basic idea is that we first construct a smooth gauge in the bulk BZ
that is periodic only in the $k_x$ and $k_y$ directions.
Then we can extract the unitary matrix $U(k_x,k_y)$
describing the gauge discontinuity
(Eq.~(\ref{eq:statement-bloch})) by calculating the overlap matrix
between the Bloch states in the top-plane
and bottom-plane gauges. The logarithm of $U(k_x, k_y)$, taken with
a given branch choice, is the $B$ matrix. We also need to
calculate the Berry curvature and Berry connections either
in the top-plane gauge or in the bottom-plane gauge.
Then all the formulas derived in previous sections
can be applied.

To be specific, we first need to construct a smooth and periodic
gauge on an arbitrary $(k_x, k_y)$ plane. For definiteness
suppose this is the $k_z\!=\!0$ plane.
We start by constructing the
``1D maxloc" gauge (see Appendix \ref{appen:pt})
along the $k_y$ direction at $k_x\!=\!0$,
then make a set of separate parallel transports
from $k_x\!=\!0$ to $k_x\!=\!2\pi$ at each $k_y$,
leaving some gauge discontinuity at the line
$k_x\!=\!2\pi$ denoted by $Y(k_y)\!=\!e^{-iD(k_y)}$.
We then apply a local (in $\k$ space) unitary transformation
$R(k_x,k_y)\!=\!e^{ik_x D(k_y)/2\pi}$ to the
occupied states at each point in the 2D plane
to smooth out this discontinuity.
In the above operation, we have maintained the smoothness
of the gauge because the $R$ matrix is defined so as to
be smooth in the interior of the 2D plane.
Furthermore, the
gauge discontinuity at the boundary line $k_x\!=\!2\pi$ has been removed.
After these operations, we have successfully constructed a
smooth and periodic gauge in the chosen $k_z\!=\!0$ plane.

Taking this gauge in the $k_z\!=\!0$ plane
as a ``reference gauge," at each $(k_x$, $k_y)$ we further
carry out two sets of parallel transports
along the positive and negative $k_z$ directions from $k_z\!=\!0$ to
$k_z\!=\!\pm\pi$. However, now the periodicity condition in $k_z$
is relaxed so that the states are as aligned to
each other as possible in the interval $k_z\!\in\!(-\pi,\pi)$. This makes the
numeric convergence of the bulk integral, \eq{theta-b}, much easier.
The overall result is a gauge that is smooth
everywhere in the bulk BZ and periodic only in
the $k_x$ and $k_y$ directions.
Some gauge discontinuity is left at the plane $k_z\!=\!\pm\pi$,
which is described by the $U$ matrix introduced in Sec.~\ref{sec:statement}.
We are now prepared to apply the formulas derived in
Sec.~\ref{sec:surf} and Sec.~\ref{sec:vl} to our system of interest.

The above procedures have to be implemented with
caution if the system is
in the \Z2-odd\ phase. In this case, it is
desirable to construct a bulk gauge respecting
$\cT$ symmetry, so that both $\theta\BK$ and
$\theta\GD$ vanish, and the remaining
contribution from $\theta\VL$ is quantized as $\pm\pi$.
For a 3D strong TI, however, the 2D \Z2\ indices for
the $k_z\!=\!0$ plane and the $k_z\!=\!\pi$ plane must be opposite.
Since it is impossible to construct
a smooth and periodic $\cT$-symmetric gauge
in the \Z2-odd\ plane,\cite{fu-prb06}
one has to select the \Z2-even\ plane for the construction of the
reference gauge.
Since standard methods for computing \Z2\ indices are now
available,\cite{soluyanov-prb11,dai-prb11} even in the absence of
$\cP$ symmetry, the selection of the \Z2-even\ plane should be
straightforward.

\begin{figure}
\centering
\includegraphics[width=8cm]{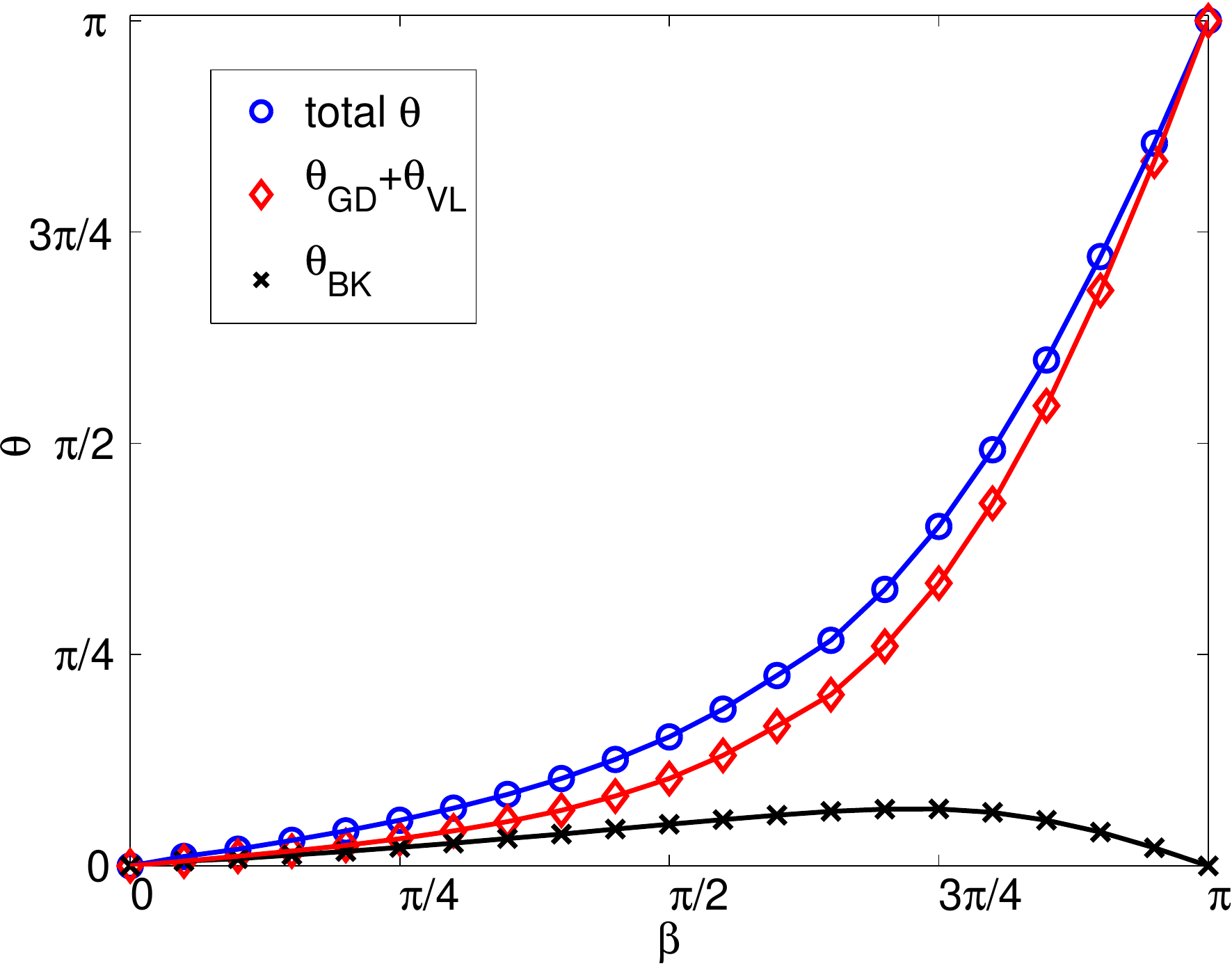}
\caption{Axion response $\theta$ for the Fu-Kane-Mele model.
Blue circles denote the total response.
Red diamonds indicate the contribution
from the gauge discontinuity, including both
the 2D surface integral $\theta\GD$
and the 1D vortex-loop integral $\theta\VL$.
Black crosses represent the
contribution from the bulk integral without
enforcing periodicity in the $k_z$ direction. }
\label{fig:theta}
\end{figure}

The axion response $\theta$ for the FKM model is shown
as blue circles in Fig.~{\ref{fig:theta}}.  As $\beta$
increases from 0 to $\pi$, the system evolves from
a \Z2-even\ to a \Z2-odd\ phase without
closing the bulk energy gap, and
$\theta$ increases smoothly from $0$ to $\pi$. 
When $\beta$ is below $\sim\!0.85\pi$,
a conventional 3D numeric integral using a fully smooth and 
periodic gauge throughout the BZ is still practical, and 
the results obtained from our method are perfectly consistent
with those from the conventional method in this regime. 
Nevertheless, it is much easier to reach numerical convergence
using our method. For example,
when $\beta\!=\!0.85\pi$,  the conventional 
method requires a $120\times 120\times 120$ 
$\k$ mesh to reduce the numerical error to within $1\%$, while
only an $80\times 80\times 80$ $\k$ mesh  
is needed to obtain the same numerical convergence
using our method.
When $\beta$ exceeds $0.85\pi$, 
it becomes impractical to get the expected convergence using 
the conventional method, 
and the advantage of our method becomes more obvious.
For example,
when $\beta\!=\!0.9\pi$, the bulk integral using
the conventional method (enforcing periodicity in all three directions)
does not converge to the expected value 
even for a $200\!\times\!200\!\times\!200$ $\k$ mesh,
\footnote{
  The convergence of $\theta$ using the conventional method 
  is trapped into some local minimum when
  $\beta\!>\!0.85\pi$. For example, when $\beta\!=\!0.9\pi$, the converged value for
  $\theta$ with a $200\!\times\!200\!\times\!200$ mesh is 0.819, 
  which is about 38.5\% of the value obtained from our method.}
while it 
converges easily for a $100\!\times\!100\!\times\!80$ $\k$ mesh 
for $\theta\BK$ in our method.
The 2D integral $\theta\GD$ 
also converges with a $100\!\times\!100$ 2D $\k$
mesh after the bulk gauge is constructed. 
The convergence for the vortex-loop integral ($\theta\VL$) is even easier;
discretizing the loop into $\sim$40 $\k$ points would typically be 
enough to get converged values of 
Berry phases (a $100\!\times\!100$ 2D $\k$
mesh discretizes the vortex loop into 41 $\k$ points when $\beta\!=\!0.9\pi$
with the branch choice $(-7\pi/4, \pi/4]$). 
Summing over all three terms $\theta\BK$, $\theta\GD$
and $\theta\VL$ eventually leads to the results indicated
by blue circles in Fig.~\ref{fig:theta}.

Note that the axion coupling of the FKM model has been calculated
previously using other methods.
In Ref.~\onlinecite{essin-prl09}, when $\beta$ approaches $\pi$,
Essin \textit{et al.}\ switched to some indirect methods such as calculating
the total polarization of a finite sample subject to a weak
external magnetic field;
while in Ref.~\onlinecite{maryam-prl15} Taherinejad \textit{et al.}\ calculated
$\theta$ in the ``hybrid-Wannier-function" basis. The results obtained from 
our method also agree very well with
these previous results when $\beta$ is close to $\pi$.

As shown in Fig.~{\ref{fig:theta}}, 
it is helpful to decompose the total $\theta$ into the
bulk-BZ integral $\theta\BK$ and
the remainder $\theta\GD+\theta\VL$, which
are indicated by black crosses and red diamonds respectively.
One finds that as $\beta$ increases,
$\theta\GD+\theta\VL$ becomes more and more dominant.
Eventually when $\beta\!=\!\pi$, $\theta$ comes entirely from
by the vortex-loop term, which equals $\pi$,
because both $\theta\GD$ and $\theta\BK$
vanish due to the $\cT$-symmetric bulk gauge.

It should be noted that none of the three terms
$\theta\BK$, $\theta\GD$ or $\theta\VL$,
is independently gauge invariant.
As the size of the vortex loop is dependent on the branch choice,
in general both $\theta\VL$ and $\theta\GD$
are branch-choice dependent, but the sum of them
should remain invariant if the bulk gauge is fixed.
\begin{figure}
\centering
\includegraphics[width=8cm]{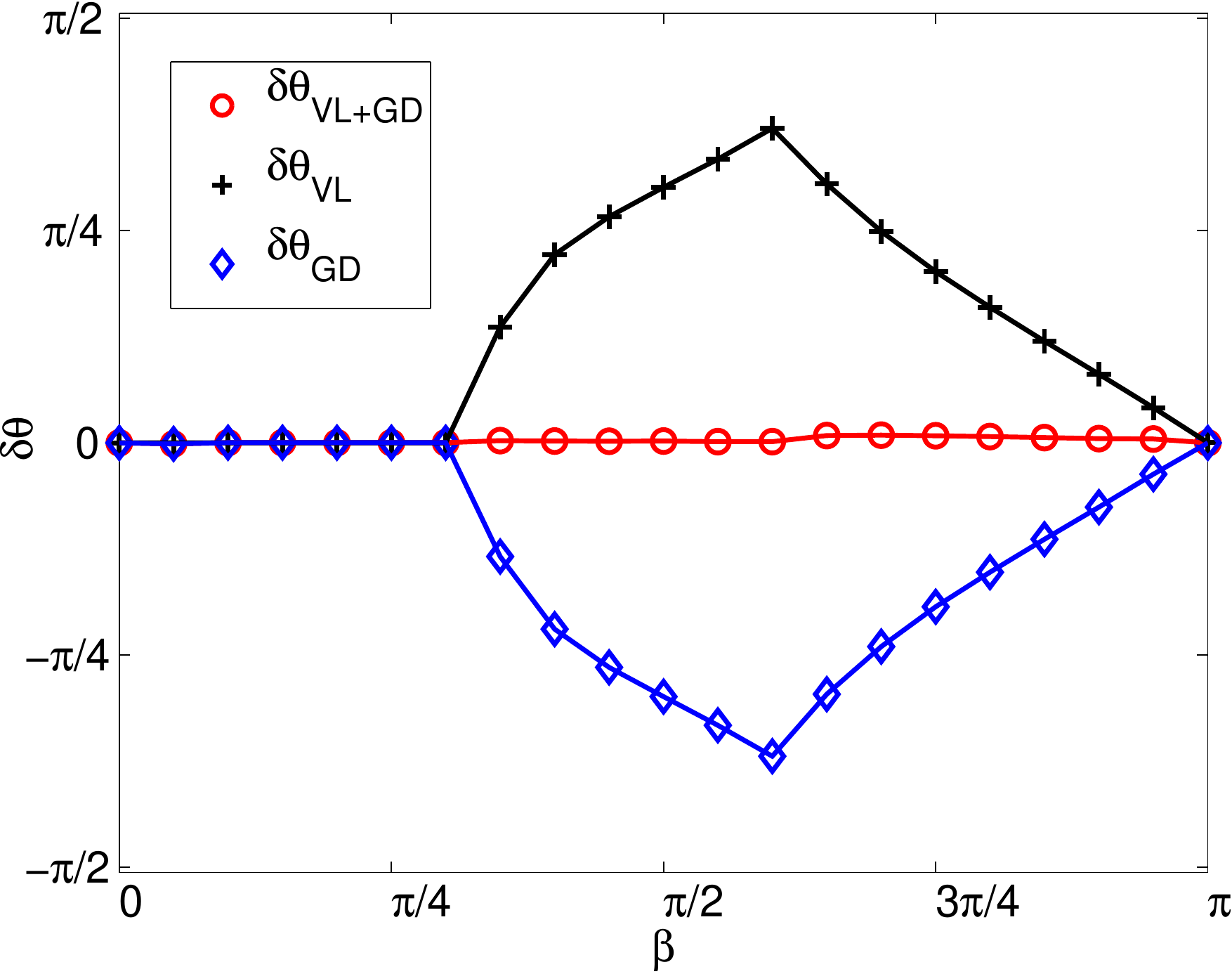}
\caption{Difference between the $\theta$ values
calculated with two different branch choices (see text) for the
Fu-Kane-Mele model.
The blue diamonds, black pluses and red circles
denote the differences for
$\theta\GD$, $\theta\VL$ and $\theta\VL+\theta\GD$
respectively.}
\label{fig:branchtest}
\end{figure}

The above statement is verified
by computing $\theta\VL$ and $\theta\GD$ using
different branch choices for a given gauge in the bulk BZ
as shown in Fig.~\ref{fig:branchtest}, where
the blue diamonds (black plus signs) denote the difference
between the values of $\theta\GD$ ($\theta\VL$)
calculated using the two different branch choices
$(-2\pi, 0]$ and $(-7\pi/4, \pi/4]$.
For the first branch choice $(-2\pi, 0]$,
a vortex loop appears when $\beta\!=\!0.35\pi$ and then
grows as $\beta$ increases, while
for the other branch choice the $B$ matrix remains
continuous throughout the 2D $\k$ plane until $\beta\!=\!0.65\pi$.
It is clearly seen from Fig.~\ref{fig:branchtest} that
both $\theta\VL$ and $\theta\GD$ depend on the branch choice.
On the other hand,  the red circles in Fig.~\ref{fig:branchtest} represent
the difference of the total $\theta\GD+\theta\VL$ computed for the
two different branch choices.  The difference
remains vanishingly small throughout the adiabatic path, thus
numerically confirming
that the sum of $\theta\VL$ and $\theta\GD$
remains branch-choice-invariant.

Besides the branch choice,
there is still the freedom to choose the gauge in the bulk BZ;
both $\theta\BK$ and $\theta\GD +\theta\VL$ depend on this
gauge choice. However,
since the bulk gauge was chosen in such a way as to align the states
with each other as much as possible
in the $k_z$ direction, the bulk integral $\theta\BK$
is typically small, explaining
why $\theta\GD+\theta\VL$ dominates over $\theta\BK$ in
Fig.~\ref{fig:theta}.

%---------------------------------------------------------------------------
\section{Summary}
\label{sec:summary}
%---------------------------------------------------------------------------

To summarize, we have developed a new method for computing the
Chern-Simons axion coupling $\theta$.
The basic idea is to relax the periodicity condition
of the gauge in one of the $\k$ directions,
thus introducing a gauge discontinuity
residing at a 2D \k\ plane in the BZ.
The total $\theta$ then has both a bulk contribution $\theta\BK$,
obtained as a conventional 3D integral over the interior of the
bulk BZ, and a gauge-discontinuity contribution $\theta\GD$,
which is expressed as a 2D integral over the gauge-discontinuity
plane as given by Eqs.~(\ref{eq:tGD}) and (\ref{eq:Gfull}).
Moreover, it may happen that discontinuities are introduced
for a given branch 
choice of $B(\k)$, the logarithm of the unitary
connection matrix describing the gauge discontinuity; this is
sometimes done for convenience, but may also be required depending
on the topological properties of the system.  In such cases the
gauge-discontinuity plane is further divided into subregions
by 1D vortex loops, and one must also consider the vortex-loop
contribution as expressed in \eq{vl_final}.  The total $\theta$
is then  $\theta\!=\!\theta\BK+\theta\GD+\theta\VL$.

Since the periodicity condition in one of the \k\
directions (e.g., the $k_z$ direction) is relaxed,
the gauge in the bulk BZ does not twist as strongly as in the case when both
periodicity and smoothness are required. This leads to improved
numerical convergence of the 3D bulk integral of \eq{theta-b}.
The loss of periodicity is compensated by extra
contributions from the gauge discontinuity ($\theta\GD$) and possible
vortex loops ($\theta\VL$). The formulas for both terms
turn out to be fairly simple and can be implemented
numerically without difficulty.

It is interesting to note that if a $\cT$-respecting gauge
has been constructed in the bulk BZ for a $\cT$-invariant system,
then both $\theta\BK$ and $\theta\GD$ must vanish. The
only surviving term $\theta\VL$ is then either
0 or $\pi$, corresponding to the \Z2\ classification of 3D $\cT$-invariant
insulators. Our theory thus provides a new
interpretation to the formally quantized magnetoelectric response in TIs.

We have applied our method to the Fu-Kane-Mele model with
staggered Zeeman field. We calculated
the axion response for the model
along a $\cT$-breaking path connecting
the \Z2-even\ and \Z2-odd\ phases.
Our results agree well with the previous
results obtained from other methods.\cite{essin-prl09,maryam-prl15}
In particular, we find that the
gauge-discontinuity contribution $\theta\GD+\theta\VL$ becomes
increasingly dominant as the system approaches the \Z2-odd\ phase.
In the TI phase, as mentioned above,
$\theta$ is completely determined by the
vortex-loop term for a $\cT$-symmetric
gauge in the bulk BZ, and  the $\pi$
quantization of $\theta$ is due to the $\pi$
quantization of the Berry
phase around a single vortex loop.

Our method may be generalized to the case that
the 3D BZ is divided into multiple subvolumes.
These subvolumes may meet each other at multiple 2D surface patches,
each with its own gauge discontinuity.
The surface patches may further meet at 1D lines or curves, which
may be vortex loops.
In such more complicated cases, the formula for $\theta\GD$
still applies, but the definition of a vortex line
has to be generalized to the situation
that the $U$ matrices obtained by approaching the vortex loop
from different surface patches may  no longer
commute with each other. Thus the formula for $\theta\VL$ may need to
be modified. We leave this problem for future study.

 From a theoretical point of view, the results presented in
this paper provide a step forward in understanding
the axion coupling in 3D insulators.
We introduced the gauge-discontinuity
and vortex-loop contributions to $\theta$,
and found that the latter can be
expressed in an unusual way as the sum of two closely
related Berry phases. From the perspective of
first-principles calculations, we have proposed a numerically
efficient method for computing the Chern-Simons orbital ME coupling
in solids. Our method can be implemented straightforwardly in standard
first-principles code packages. This makes
it possible to compute the orbital ME coefficients efficiently
for realistic materials, thus facilitating the search for
functional materials with enhanced ME couplings.

\acknowledgments

This work was supported by NSF Grant DMR-14-08838.

\appendix
%--------------------------------------------------------
\section{Parallel transport}
\label{appen:pt}
%--------------------------------------------------------

In this section, 
we discuss how to carry out the parallel transport operation
and construct the ``1D maxloc" gauge starting from
a set of occupied eigenstates with an arbitrary
gauge on a given \k\ path. The basic idea is to
recursively make the (periodic part of)
the Bloch states at each $\k$ point on the path as aligned
as possible with the states at the immediately previous $\k$ point.
If the $\k$ path is chosen to be a closed
loop, the states
obtained at the end of the loop may differ from those at
the start by some phase factors
after the parallel-transport operation, with the mismatch
giving exactly the Berry phases of 
the Bloch states around the loop. The Berry phases accumulated along 
the $\k$ path can be smeared out
by smoothly distributing the phase to the states 
at every $\k$ point on the loop. After such an
operation, one obtains a gauge which is both smooth and periodic along
the path (loop), and which we refer to as a
1D maxloc gauge.\cite{MLWF-1}

To be specific, let us consider a set of occupied bands
$\ket{u_{n\k}}$, with $n=1, ... N$,
which are isolated from other bands (in energy) everywhere in the BZ.
Let us take a closed $\k$ path along $k_z$ running from 0 to $2\pi$,
with the path sampled by $J$ discrete points,
so that $\k_j=2\pi(j-1)\hat{z} /J $.
Assume that the eigenstates with some arbitrary
gauge $\ket{u_{n\k_j}^{0}}$
are given for $j=1,...J$, and a periodic gauge is chosen
at the $(J+1)$th point so that
$\ket{u_{n\k_{J+1}}^0}=e^{-i2\pi z}\ket{u_{n\k_{1}}^0} $.
To carry out the parallel transport, we need to insist that
each overlap matrix between the occupied states at $\k_{j+1}$
and $\k_{j}$, i.e., $M_{mn}(j)={\ip{u_{m\k_{j}}^0}{u_{n\k_{j+1}}^0}}$,
is positive-definite Hermitian.
This can be done as follows.
At each $\k_{j}$, make a singular-value decomposition to the overlap
matrix $M_j=V_j\Sigma_j W\dag_j$, where
$V$ and $W$ are unitary and $\Sigma$ is real positive diagonal.
Then apply a unitary transformation
$L_j=W_jV\dag_j$ to $\ket{u_{n\k_{j+1}}^0}$,
$\ket{\widetilde{u}_{n \k_{j+1}}}=\sum_{m=1}^{N} L_{j,mn}\ket{u_{m\k_{j+1}}^0}$,
so that the overlap matrix between the unitarily transformed states at neighboring
$\k$ points becomes positive-definite Hermitian.
If one repeats such an operation from $j=1$ to $j=J$, the states will become
as aligned to each other as possible at all $\k$ points on the path.
However, there is still some gauge discontinuity left at the boundary,
$\ket{\widetilde{u}_{n\k_{J+1}}}=
e^{-i2\pi z} \sum_{m} \Lambda_{mn}\ket{\widetilde{u}_{m\k_{1}}}$,
where $\Lambda$ is a unitary matrix. The logarithms of the
eigenvalues of $\Lambda$, $\beta_{n}=-i\ln \lambda_{n}$,
are then identified as the non-Abelian Berry phases (also known as
Wilson loop eigenvalues)
of the Bloch states.

The gauge obtained from the parallel-transport operation 
is smooth along the $\k$ path,
but not periodic. To restore the periodicity, we need to rotate all the states on
the $\k$ path to the basis that diagonalizes $\Lambda$, i.e.,
$\ket{u'_{n\k_{j}}}=\sum_{m=1}^{N}\ket{\widetilde{u}_{m\k_{j}}}\,L_{mn}$, where $L$
is the eigenvector matrix of $\Lambda$.  Then we gradually smear out the discontinuity
by applying the following phase twist to the states at $\k_{j}$:
$\ket{u_{n\k_{j}}}=e^{-i(j-1)\beta_n/J}\ket{u'_{n\k_{j}}}$.
This results in a ``1D maxloc gauge'' that is both smooth
and periodic along the $\k$ path.

%-------------------------------------------------------------
\section{Integration over $\lam$ in the formula for $\theta\GD$}
\label{appen:int}
%------------------------------------------------------------

In deriving the expression in \eq{tGD} for the gauge discontinuity
contribution $\theta\GD$ in Sec.~\ref{sec:formalism}, we arrived
at \eq{Gfull} involving the quantities $\Bbbar_x$, $\Bbbar_y$, and
$\Bbbar_\cxy$ which were expressed as integrals over $\lambda$
in Eqs.~(\ref{eq:bbx}-\ref{eq:bbxy}).  We show here that these
three quantities can all be computed analytically in the sense
that the $\lambda$ integral does not have to be discretized.

The plan is as follows.
Suppose that $\Ao_x(\k)$, $\Ao_y(\k)$, $\O^{(0)}_{xy}(\k)$,
$B(\k)$, $B_x(\k)$, and $B_y(\k)$ are known.  
The first term in \eq{Gfull} is independent of $\lam$ and is
trivial.  For the remaining terms, at each $\k$, locally diagonalize $B(\k)$,
then transform all of the matrices $\Ao_x(\k)$, $\Ao_y(\k)$,
$B_x(\k)$, and $B_y(\k)$ to the basis that locally diagonalizes $B(\k)$,
i.e.,
\bea
&&B(\k)\to V\dag(\k)\,B(\k)\,V(\k), \;\nn
&&B_{a}(\k)\to V\dag(\k)\,B_{a}(\k)\,V(\k), \;\nn
&&A_{a}(\k)\to V\dag(\k)\,A_{a}(\k)\,V(\k), \;\nn
&&\O_{xy}(\k)\to V\dag(\k)\,\O_{xy}(\k)\,V(\k)\; ,
\eea
where $V(\k)$ is the eigenvector matrix of $B(\k)$, and
$a=\{x, y\}$.
Then one can compute
the trace in this basis.
Letting $B_{mn}=b_n\,\delta_{mn}$, we find
\bea
\Bbar_{x,mn}\ol &=&\int_0^\lam d\mu\;
e^{-i\mu b_m}\,B_{x,mn}\,e^{i\mu b_n}\;\nn
&=&g_{mn}\ol\,B_{x,mn} \; ,
\eea
where
\beq
g_{mn}\ol=\frac{e^{-i\lambda(b_m-b_n)}-1}{-i(b_m-b_n)} \;.
\eeq
Then
\bea
\Bbbar_{x,mn}&=&\Big(\int_0^1 g_{mn}(\lam) \, d\lam \Big) \; B_{x,mn}\;\nn
&=&\Big(\frac{e^{i(b_n-b_m)}-1}{-(b_n-b_m)^2}-\frac{1}{i(b_n-b_m)}\Big)\;B_{x,mn}\;\nn
\label{eq:Bbbar_x_anly}
\eea
and
\bea
\Bbbar_{\cxy mn}=&&i \sum_l
\Big(\int_0^1 g_{ml}(\lam)\,g_{ln}(\lam) \, d\lam \Big)\,\times \;\nn
&&\Big( B_{x,ml} \, B_{y,ln} - B_{y,ml} \, B_{x,ln} \Big) \;.
\label{eq:Bbbar_xy}
\eea
Because we are interested in the trace of $B\,\Bbbar_\cxy$ in the basis that
$B$ is locally diagonal,  only the diagonal matrix elements of $\Bbbar_\cxy$
are relevant. After carrying out the integral in \eq{Bbbar_xy} one obtains the
following expression:
\bea
\Bbbar_{\cxy nn}=&&i\sum_{m}
\Big(\frac{2}{(b_n-b_m)^2}-\frac{2\sin(b_m-b_n)}{(b_m-b_n)^3}\Big)\times \;\nn
&&\Big(B_{x,nm}B_{y,mn}-B_{y,nm}B_{x,mn}\Big)\;.
\label{eq:Bbbar_xy_anly}
\eea
If two eigenvalues $b_m$ and $b_n$ are degenerate, one needs to take the limit
$(b_n-b_m)\to 0$. It turns out that both quantities are finite:
\beq
\lim_{b_n\to b_m}\Bbbar_{x,mn}=B_{x,mn}/2\; ,
\eeq
and
\beq
\lim_{b_m\to b_n}\Bbbar_{\cxy nn}=
 \frac{i}{3}\Big(B_{x,nm}B_{y,mn}-B_{y,nm}B_{x,mn}\Big)\; .
\eeq

Of course the entire calculation still has to be done on
a discretized mesh on the $\k$ plane, with finite-difference
expressions used to evaluate objects like $\Ao_x(\k)$, so it
it not ``exact". However, it is convenient that we don't have
to discretize the $\lam$ axis, instead doing all $\lam$
integrals analytically.

%--------------------------------------------------------------
\section{Derivation of \eq{Atrsym}}
\label{appen:trsym}
%--------------------------------------------------------------

In Sec.~\ref{sec:TR_tGD} we considered the effect of
time-reversal symmetry on the gauge discontinuity on the
boundary plane and showed that $\theta\GD$ has to vanish
for a $\cT$-respecting choice of gauge.  The demonstration
rested on the use of \eq{Atrsym}, which was only introduced
heuristically there.

Here we prove it properly.  From
Eqs.~(\ref{eq:Alxy}), (\ref{eq:cAxy}) and (\ref{eq:Bbarxy}),
we know that
\beq
 A_{a}^{(\lam)}
=\widetilde{A}_{a}^{(\lam)}+\Gamma_{a}(0,\lam)
\label{eq:Axylam}
\eeq
where
\beq
\widetilde{A}_{a}^{(\lam)}=W\dag(\lam)\,A_{a}^{(0)}\,W(\lam) \; ,
\label{eq:Atxy}
\eeq
and the function $\Gamma_{a}(\lam_1,\lam_2)$ is defined as
\beq
\Gamma_{a}(\lambda_1, \lambda_2)=\int_{\lambda_1}^{\lambda_2}d\mu\,
W\dag(\mu)\,B_{a}\,W(\mu)\; .
\label{eq:Gammaxy}
\eeq
Letting
$\lam=1$, we get the expression
\beq
 A_{a}^{(1)}
=\widetilde{A}_{a}^{(1)}+\Gamma_{a}(0,1),
\label{eq:A1}
\eeq
Applying a unitary transformation $W(1-\lam)$ to the
matrix $A_{a}^{(1)}$, one obtains 
\bea
W(1-\lam)\,A_{a}^{(1)}\,W\dag(1-\lam)
&=& \widetilde{A}_{a}^{(\lam)}+\Gamma_{a}(\lam-1, \lam) \; ,\nn
&=& A_{a}^{(\lam)}+\Gamma_{a}(\lam-1,0)\;,\nn
\label{eq:wA1w}
\eea
where a variable transformation $(\lam + \nu - 1) \!\to\! \mu $
has been made to obtain the second term $\Gamma_{a}(\lam-1, \lam)$
on the RHS of the first line in \eq{wA1w}.
The integral from $\lam-1$ to $\lam $ in $\Gamma_{a}(\lam-1,\lam)$
is further divided into  two integrals: one from $\lam-1$  to 0,
and the other from 0 to $\lam$.
$A_{a}^{(\lam)}$ in the second line is then obtained
by combining the integral
from $0$ to $\lam$ together with
$W\dag(\lam)\,A_{a}^{(0)}\,W(\lam)$ (\eq{Axylam}).
Therefore
\beq
A_{a}^{(\lam)}=W(1-\lam)\,A_{a}^{(1)}\,W\dag(1-\lam)\;
-\Gamma_{a}(\lam-1,0)
\eeq
and it immediately follows that
\bea
A_{a}^{(1-\lam)}&=&W(\lam)\,A_{a}^{(1)}\,W\dag(\lam)-
\Gamma_{a}(-\lam, 0)\;\nn
&=&W(\lam)\,A_{a}^{(1)}\,W\dag(\lam)-
\int_{0}^{\lam}d\mu\,W(\mu)\,B_{a}\,W\dag(\mu)\; , \nn
\label{eq:A_1-l}
\eea
where we let $\mu\to -\mu$ in going from the first to the second line in
\eq{A_1-l}. Equation~(\ref{eq:Atrsym} then follows by combining
Eq.~(\ref{eq:trsym2})-(\ref{eq:trsym3}) and \eq{A_1-l}:
\begin{widetext}
\bea
A_{a}^{(1-\lam)}(-\k)\;
&=&e^{-i\lam B(-\k)}\,A_{a}^{(1)}(-\k)\,e^{i\lam B(-\k)}-
\int_{0}^{\lam}d\mu\,e^{-i\mu B(-\k)}\,B_{a}(-\k)\,e^{i\mu B(-\k)}\;\nn
&=&\sigma_y\,e^{-i\lam B^{T}(\k)}\,\Big(A_{a}^{(0)}(\k)\Big)^{T}\,
e^{i\lam B^{T}(\k)}\,\sigma_y
+\sigma_y\,\int_{0}^{\lam}d\mu\,
e^{-i\mu B^{T}(\k)}\,B_{a}^{T}(\k)\,e^{i\mu B^{T}(\k)}\,\sigma_y\;\nn
&=&\sigma_y\,\Big(W\dag(\lam)\,A_{a}^{(0)}(\k)\,W(\lam)+
\int_{0}^{\lam}d\mu\,W\dag(\mu)\,B_{a}(\k)
\,W(\mu)\Big)^{T}\,\sigma_y\;.
%&&=\sigma_y\,\Big(A_{x(y)}^{(\lam)}(\k)\Big)^{T}\,\sigma_y \; .
\label{eq:A_1-l_long}
\eea
\end{widetext}
The last line in \eq{A_1-l_long} is simply
$\sigma_y\,\Big(A_{a}^{(\lam)}(\k)\Big)^{T}\,\sigma_y$,
thus proving \eq{Atrsym} and thereby confirming that $\theta\GD$
vanishes for a TR-invariant gauge.

%--------------------------------------------------------------
\section{Derivation of \eq{vl1}}
\label{appen:vl}
%--------------------------------------------------------------

In Sec.~\ref{sec:vl-derive} we proposed a formula for the
vortex-loop contribution as expressed in \eq{vl1}. 
We only explained the main idea there, and the formula was introduced
without proof. Here we provide a rigorous derivation.

To derive \eq{vl1},
it is convenient to decompose $G(\k)$ into
four terms $G_1$, $G_2$, $G_3$ and $G_4$ corresponding to
the four terms on the right-hand side (RHS) of \eq{Gfull}:
\bea
&&G_1=B\,\O^{(0)}_{xy} \;,
\label{eq:g1}\\
&&G_2=iB\,[\,\Bbar_x(\lambda),\Bbar_y(\lambda)\,] \;,
\label{eq:g2}\\
&&G_3=iB\,[\,\Ao_x, \Bbar_y(\lambda)\,]\;,
\label{eq:g3}\\
&&G_4=iB\,[\,\Bbar_x(\lambda),\Ao_y\,] \;.
\label{eq:g4}
\eea
Since all the quantities such as $\O_{xy}$ and $A_{x(y)}$ are defined
in the bottom-plane gauge, we will drop the superscript
``(0)" (indicating the bottom-plane gauge) in later steps.
Recalling that the change $\Delta B$ in the interior region
was expressed in \eq{dB} as $V\Delta_1\V^\dagger$, where $V$
is the unitary matrix that diagonalizes $B$ and $\Delta$ is diagonal
with $2\pi$-integer entries, we can transform the needed matrices
to the $B$-diagonal representation via
\bea
&&A_{a}'=V\dag\,A_{a}\,V \; ,
\label{eq:Axyprime}\\
&&\O_{xy}'=V\dag\,\O_{xy}\,V\; ,
\label{eq:Oxyprime}\\
&&\Bbar_{a}'=V\dag\,\Bbar_{a}\,V \; .
\label{eq:Bbarprime}
\eea
We will prove \eq{vl1}
by explicitly calculating the four terms in
Eqs.~(\ref{eq:g1})-(\ref{eq:g4}).

%%%%%%%%%%%%%%%%%%%%%%%%%%%%%%%%%%%%%%%%%%%%%%%%%%%%%%%%%%%%%%%%%%%
\subsection{ The $G_1$ term}
%%%%%%%%%%%%%%%%%%%%%%%%%%%%%%%%%%%%%%%%%%%%%%%%%%%%%%%%%%%%%%%%%%%%%%

Plugging Eq.~(\ref{eq:dB}) first into the expression
for $G_1$ in \eq{g1}, one obtains
\bea
\Tr\Big[\,G_1\,\Big]&=&\Tr\Big[\,V\,\Delta_1\,
V^{\dagger}\O_{xy}\,V\,V^{\dagger}\,\Big]\;\nn
&=&\Tr\Big[\,\Delta_1\,V^{\dagger}\O_{xy}\,V\,\Big]\;.
\eea
Note that $\O_{xy}'=V\dag \,\O_{xy}\,V$ is associated
with the Berry curvature of the Bloch states in the bottom-plane gauge
that are unitarily transformed by $V$:
$\ket{\bar{u}^{(0)}_n}=\sum_{m=1}^{N}\ket{u^{(0)}_m}\,V_{mn}$.
One can express the Berry curvature of $\ket{\bar{u}^{(0)}_n}$
(denoted by $\bar{\O}_{xy}$) in terms of
$A_x$, $A_y$, $\O_{xy}$, $V$ and
the partial derivatives of $V$,
\beq
\bar{\O}_{xy}=\O_{xy}'+\Lambda_{xy}+
i[\,C_x,  A_y'\,]-i[\,C_y, A_x'\,]\; ,
\label{eq:oprime}
\eeq
where $C_x$ and $C_y$ are defined in \eq{Cxy},
and
\beq
\Lambda_{xy}=\partial_x C_y -\partial_y C_x \;\,
\label{eq:Lamxy}
\eeq
can be considered as the Berry curvature
in the ``gauge space."

 From Eq.~(\ref{eq:oprime}) it immediately follows that
\bea
\Tr\Big[\,G_1\,\Big]=&&\Tr\Big[\,\Delta_1\,\bar{\O}_{xy}-
\Delta_1\,\Lambda_{xy}-i\Delta_1\,[\,C_x , A_y' \,]\;\nn
&&+i\Delta_1\,[\,C_y,  A_x'\,]\,\Big] \;.
\label{eq:g1_long}
\eea
Before further simplifying Eq.~(\ref{eq:g1_long}), let us go to
the other terms and come back to $G_1$ later.

\subsection{The $G_3$ and $G_4$ terms}

Let us deal with the $G_3$ and $G_4$ terms.
Since $B_x$ and $B_y$ are involved in $G_3$ and $G_4$,
let us first evaluate these two terms.
\bea
B_x &=&\partial_{x} (\,V\,\Delta_1\,V\dag\,) \;\nn
&=&\partial_x V\,\Delta_1\,V\dag+  V\,\Delta_1\,\partial_x V\dag \;\nn
&=&iV\,[\,\Delta_1 , C_x\,]\,V\dag \;.
\label{eq:bx}
\eea
Similarly, $B_y=iV\,[\,\Delta_1 , C_y\,]\,V\dag$.
Plugging the expressions for $B_x$ and $B_y$ into
Eq.~(\ref{eq:Bbarxy}),
one immediately obtains
\bea
&& \Bbar_{a}(\lam)=\int_{0}^{\lam}\,du \,
V\,e^{-iu\Delta_1}\,i[\,\Delta_1 , C_{a}\,]\,e^{iu\Delta_1}\,V\dag
\; . \nn
\label{eq:bbar_long}
\eea

We now evaluate $\Tr[G_3]$ by carrying out the
trace in the basis that diagonalizes $B$ using
Eqs.~(\ref{eq:Axyprime}-\ref{eq:Bbarprime}).  We find
\bea
\Tr\Big[\,G_3\,\Big]&=&
\Tr\Big[\,i\D_1\,[\,A_x' , \, \Bbar_y',] \Big] \;\nn
&=&\Tr\Big[\,\int_{0}^{\lam}du\,i\D_1\,[\,A_x' ,
\,e^{-iu\Delta_1}\,i[\,\Delta_1 , C_{y}\,]\,e^{iu\Delta_1}\,] \,\Big] \;\nn
&=&\Tr\Big[\,\int_{0}^{\lam}du\, i A_x' \, [\,e^{-iu\Delta_1}\,i[\,\Delta_1 , C_{y}\,]
\,e^{iu\Delta_1},\, \D_1\,] \,\Big] \;\nn
&=&\Tr\Big[\,i A_x' \,\int_{0}^{\lam}du\,\partial_u
\Big(\,e^{-iu\D_1}\,[\,\D_1,C_y\,]\, e^{iu\D_1}\Big)\,\Big] \;\nn
&=&\Tr\Big[\,i A_{x}'\,e^{-i\lam\D_1}\,[\,\D_1 , C_y\,]\,e^{i\lam\D_1}-
i A_x' \,[\,\D_1 , C_y\,]\,\Big] \; ,\nn
\label{eq:g3_long}
\eea
where we have used the equation
\beq
 [\,e^{-iu\Delta_1}\,i[\,\Delta_1 , C_{y}\,]
\,e^{iu\Delta_1},\, \D_1\,]
=\partial_u\Big(\,e^{-iu\D_1}\,[\,\D_1,C_y\,]\, e^{iu\D_1}\Big)
\label{eq:identity1}
\eeq
when going from the third to the fourth line in \eq{g3_long}.
Making use of the cyclic property of trace,
one immediately realizes that the second term in the
last line of Eq.~(\ref{eq:g3_long}) cancels the last
term on the RHS of Eq.~(\ref{eq:g1_long}),
which will be dropped in later steps. Therefore,
\bea
\int_{0}^{1}d\lam\,\Tr\Big[\,G_3\,\Big]&=&
\int_{0}^{1}d\lam\,\Tr\Big[\,i A_x'\,e^{-i\lam\D_1}\,[\,\D_1 , C_y\,]\,e^{i\lam\D_1}\,\Big] \;\nn
&=&\int_{0}^{1}d\lam\,\Tr\Big[\,-A_x'\,\partial_{\lam}(\,e^{-i\lam\D_1}\,C_y\,e^{i\lam\D_1}\,)
\,\Big]\;\nn
&=&\Tr\Big[\,-A_x'\,(\,e^{-i\lam\D_1}\,C_y\,e^{i\lam\D_1}\,)\vert_{\lam=0}^{\lam=1}\,\Big]\;\nn
&=&0 \; ,
\label{eq:g3_cancel}
\eea
where the following equation has been used to go from the second to the third line
in \eq{g3_cancel}:
\beq
i e^{-i\lam\D_1}\,[\,\D_1 , C_y\,]\,e^{i\lam\D_1}=
-\partial_{\lam}(\,e^{-i\lam\D_1}\,C_y\,e^{i\lam\D_1}\,) \; .
\label{eq:identity2}
\eeq
Similar derivations can be applied to the $G_4$ term, i.e.,
\bea
\Tr\Big[\,G_4\,\Big]=\Tr\Big[\,i A_{y}'\,e^{-i\lam\D_1}\,[\,C_x , \D_1\,]\,e^{i\lam\D_1}-
i A_{y}'\,[\,C_x , \D_1\,]\,\Big]\;. \nn
\label{eq:g4_long}
\eea

The second term on the RHS of Eq.~(\ref{eq:g4_long})
cancels the third term on the RHS of Eq.~(\ref{eq:g1_long}).
Dropping the second term in Eq.~(\ref{eq:g4_long}) and integrating
over $\lambda$,
one obtains $\int_{0}^{1}d\lam\,\Tr\Big[\,G4\,\Big]=0$.

%------------------------------------------------------------------
\subsection{The $G_2$ term}
%------------------------------------------------------------------

In the basis that locally diagonalizes $B$,
\beq
\Tr\Big[\,G_2\,\Big]=
\Tr\Big[\,i\D_1\,[\,\Bbar_x' , \Bbar_y' \,]\,\Big] \;.
\label{eq:g2_short0}
\eeq
On the other hand, combining \eq{bbar_long}, Eq.~(\ref{eq:Bbarprime})
and \eq{identity2}, we get 
\bea
\Bbar_{a}'&=&
-\int_{0}^{\lam} d\mu
\partial_{u}(\,e^{-iu\D_1}\,C_{a}\,e^{iu\D_1} \,)\;\nn
& =& C_{a}- \widetilde{C}_{a}\;,
\eea
where $\widetilde{C}_{a}=e^{-i\lam\D_1}\,C_{a}\,e^{i\lam\D_1}$.
It follows that
\beq
\Tr\Big[\,G_2\,\Big]
 =  \Tr\Big[\,i\D_1\,[\,\widetilde{C}_x-C_x\,,\,
\widetilde{C}_y-C_y\,]\,\Big]\;.
\label{eq:g2_short}
\eeq
If one expands the RHS of \eq{g2_short},
one would obtain four commutators between $C_{a}$ and
$\widetilde{C}_{b}$ ($a, b = x, y$).
Since $e^{\pm i\lam\D_1}$ commute with $\D_1$,
the term involving $[\,\widetilde{C}_x , \widetilde{C}_y \,]$
is equal to the term with $[\,C_x , C_y\,]$.
Therefore,
\beq
\Tr\Big[\,G_2\,\Big]=
\Tr\Big[\,\D_1\,\Big(\,2i[\,C_x , C_y\,]-
i[\,C_x ,\widetilde{C}_y\,]
-i[\,\widetilde{C}_x , C_y \,]\,\Big)\,\Big]\;.
\label{eq:g2_long}
\eeq
The second term on the RHS of Eq.~(\ref{eq:g2_long})
can be written as a total derivative of $\lam$ as
\beq
\Tr\Big[-i\D_1\,[\,C_x ,\widetilde{C}_y\,]\,\Big]
%=&&\Tr\Big[-i\widetilde{C}_y\,[\D_1 , \,C_x]\,\Big]\;\nn
%=&&\Tr\Big[-i e^{-i\lam\D_1}\,C_{y}\,e^{i\lam\D_1}\,\D_1\,C_x+
%i \D_1\,e^{-i\lam\D_1}\,C_{y}\,e^{i\lam\D_1}\,C_x\,\Big]\;\nn
=-\Tr\Big[\partial_{\lam}
(\,e^{-i\lam\D_1}\,C_{y}\,e^{i\lam\D_1}\,C_x\,)\,\Big]\;.
\label{eq:g2_cancel}
\eeq
We need to use
$\D_1\,e^{\pm i\lam\D_1}=\mp i\partial_{\lam}(e^{\pm i\lam\D_1})$
to obtain the above equation.
Integrating
Eq.~(\ref{eq:g2_cancel}) over $\lam$, one obtains zero.
Similarly, after integrating over $\lam$, the third term
on the RHS of Eq.~(\ref{eq:g2_long}) also vanishes.
Therefore,
\beq
\int_{0}^{1}d\lam\,\Tr\Big[\,G_2\,\Big]
=\Tr\Big[\,2\D_1\,i[\,C_x , C_y\,] \,\Big] \;.
\eeq
Note that the gauge-covariant Berry curvature
defined in the gauge space
$\widetilde{\Lambda}_{xy}=\Lambda_{xy}-i[\,C_x , C_y\,]$
has to vanish ($\Lambda_{xy}$ defined in Eq.~(\ref{eq:Lamxy})),
because $\widetilde{\Lambda}_{xy}$
is the Berry curvature projected onto the unoccupied subspace,
which is zero. Therefore,
$\Lambda_{xy}=i[\,C_x , C_y\,]$. It can also be
shown by explicitly writing out the commutator
of $C_x$ and $C_y$:
\bea
i[\,C_x , C_y\,]\;
&=&i(-V\dag\,\partial_x V\,V\dag\partial_y V+V\dag\,\partial_y V\,V\dag\partial_x V)\;\nn
&=&i(V\dag\,V\,\partial_x V\dag\partial_y V-V\dag\,V\,\partial_y V\dag\partial_x V)\;\nn
&=&i \partial_x (V\dag\,\partial_y V)-i\partial_y (V\dag\,\partial_x V)\;\nn
&=&\Lambda_{xy} \;.
\eea
We have used the fact that $V\,V\dag=1$
and $\partial_a(V\,V\dag)=0$ in
the above derivations. Therefore,
\beq
\int_{0}^{1}d\lam\,\Tr\Big[\,G_2\,\Big]
=\Tr\Big[\,2\D_1\,\Lambda_{xy} \,\Big] \;.
\label{eq:g2_final}
\eeq

Combining Eqs.~(\ref{eq:g1_long}), (\ref{eq:g3_long}),
(\ref{eq:g4_long}) and (\ref{eq:g2_final}), we get
\bea
\theta_{\textrm{shift}}\;
&=&\frac{-1}{4\pi}\int dk_xdk_y\int_{0}^{1}
d\lam \,\Tr\Big[\,G_1+G_2+G_3+G_4\,\Big]\;\nn
&=&\frac{-1}{4\pi}\int_{\mathcal{S}} dk_x dk_y
\,\Tr\Big[\,\D_1\O_{xy}'+\D_1\Lambda_{xy}\,\Big]\;\nn
&=&\frac{-1}{4\pi}\int_{\mathcal{S}} dk_x dk_y
\,\Big(\,2\pi\,(\O_{xy}')_{11}+2\pi\,(\Lambda_{xy})_{11}\,\Big)\;\nn
&=&-\Big[\,\phi_{1}(\mathcal{C})+\xi_1(\mathcal{C})\,\Big]/2 \;.
\eea
This completes the derivation of Eq.~(\ref{eq:vl1}), demonstrating
that $\theta\VL$ is just the average of the two Berry phases
$\phi_{1}(\mathcal{C})$ and $\xi_1(\mathcal{C})$ appearing above.

\bibliography{theta}

\end{document}